\documentstyle[11pt]{article}
\newdimen\mathindent
\renewcommand{\qquad}{\hspace*{25pt}}
\newcommand{\eref}[1]{(\ref{#1})}

\newcommand{\eqll}[2]{\begin{equation}{#1}\label{#2}\end{equation}}
\newcommand{\eq}[1]{\begin{equation}{#1}\end{equation}}
\def\integer{{\mathchoice
    {\hbox{ $\displaystyle\kern-1mm {\rm Z}\kern-1.1mm {\rm Z}$}}
    {\hbox{ $\textstyle\kern-1mm {\rm Z}\kern-1.1mm {\rm Z}$}}
    {\hbox{$\scriptstyle\kern-1mm {\rm Z}\kern-1.1mm {\rm Z}$}}
    {\hbox{$\scriptscriptstyle\kern-1mm {\rm Z}\kern-1.1mm {\rm Z}$}}}}
\def\natural{{\mathchoice
    {\hbox{ $\displaystyle\kern-1.4mm 1\kern-.7mm {\rm N}$}}
    {\hbox{ $\textstyle\kern-1.4mm 1\kern-.7mm {\rm N}$}}
    {\hbox{$\scriptstyle\kern-1.4mm 1\kern-.7mm {\rm N}$}}
    {\hbox{$\scriptscriptstyle\kern-1.4mm 1\kern-.7mm {\rm N}$}}}}
\def\real{{\mathchoice
    {\hbox{$\displaystyle\kern-.2mm 1\kern-.8mm {\rm R}\kern-.2mm$}}
    {\hbox{$\textstyle\kern-.2mm 1\kern-.8mm {\rm R}\kern-.2mm$}}
    {\hbox{$\scriptstyle\kern-.2mm 1\kern-.8mm {\rm R}\kern-.2mm$}}
    {\hbox{$\scriptscriptstyle \kern-.2mm 1\kern-.8mm {\rm R}\kern-.2mm$}}}}
\def\complex{{\mathchoice
    {\hbox{$\displaystyle\kern-.2mm {\rm C}\kern-1.5mm\raise.2mm
			       \hbox{\vrule height6pt}\kern1.3mm$}}
    {\hbox{$\textstyle\kern-.2mm {\rm C}\kern-1.5mm\raise.3mm
			       \hbox{\vrule height6pt}\kern1.3mm$}}
    {\hbox{$\scriptstyle\kern-.2mm{\rm C}\kern-1.5mm\raise.2mm
			       \hbox{\vrule height3pt}\kern1.3mm$}}
    {\hbox{$\scriptscriptstyle\kern-.2mm{\rm C}\kern-1.5mm\raise.2mm
			      \hbox{\vrule height2pt}\kern1.3mm$}}}}
\def\summ#1#2#3{\sum\limits_{#1}^{#2}\lower3pt\hbox{${ }_{#3}$}}

\newtheorem{defin}{Definition}[section]

\newtheorem{theo}{Theorem}[section]

\begin{document}

\title{\raggedright
Orbit spaces of reflection groups with 2, 3 and 4 basic polynomial
invariants\footnote{This paper is partially supported by INFN and MURST
40\% and 60\%, and is carried out as part of the European Community Programme
``Gauge Theories, Applied Supersymmetry and Quantum Gravity" under contract
SCI-CT92-D789.}}

\author{G Sartori and G Valente\\
\small Dipartimento di Fisica,Universit\`a di Padova and INFN, Sezione di
Padova\\
\small I--35131 Padova, Italy (e-mail: gfsartori@padova.infn.it,
valente@padova.infn.it)}

\date{}

\maketitle

\begin{abstract}
Functions which are covariant or invariant under the transformations of a
compact linear group can be advantageously expressed in terms of functions
defined in the orbit space of the group, i.e.\ as functions of a finite set of
basic invariant polynomials.  The equalities and inequalities defining the
orbit spaces of all finite coregular real linear groups (most of which are
crystallographic groups) with at most four independent basic invariants are
determined.  For each group $G$ acting in the Euclidean space $\real^n$, the
results are obtained through the computation of a metric matrix $\widehat
P(p)$, which is defined only in terms of the scalar products between the
gradients of a set of basic polynomial invariants $p_1(x),\dots p_q(x),\
x\in\real^n$ of $G$; the semi-positivity conditions $\widehat P(p)\ge 0$ are
known to determine all the equalities and inequalities defining the orbit
space $\real^n/G$ of $G$ as a semi-algebraic variety in the space $\real^q$
spanned by the variables $p_1,\dots ,p_q$.  In a recent paper, the $\widehat
P$-matrices, for $q\le 4$, have been determined in an alternative way, as
solutions of a universal differential equation; the present paper yields a
partial, but significant, check on the correctness and completeness of these
solutions.  Our results can be easily exploited, in many physical contexts
where the study of covariant or invariant functions is important, for instance
in the determination of patterns of spontaneous symmetry breaking, in the
analysis of phase spaces and structural phase transitions (Landau's theory),
in covariant bifurcation theory, in crystal field theory and in most areas of
solid state theory where use is made of symmetry adapted functions.

\end{abstract}

\section{Introduction}
Functions which are covariant or invariant under the transformations of a
compact linear group (hereafter abbreviated in CLG) $G$ play an important role
in physics, and the determination of their properties is often a basic problem
to solve.

An example, which is relevant both to elementary particle and solid state
physics, is offered by the determination of the possible patterns of
spontaneous symmetry breaking in theories in which the ground state of the
system is determined by the minimum of an invariant potential $V(x)$.

Let us sketch the relevant physical context.  The symmetry group $G$ of the
formalism used to describe a physical system acts as a permutation group on
the set of the solutions of the evolution equations.  When the ground state of
the system is invariant only with respect to a proper subgroup $G_0\subset G$,
the $G$-symmetry is said to be {\it spontaneously broken} (see, for instance
\cite{620,051,557,558} and references therein) and $G_0$ turns out to be the
{\it true} symmetry group of the system \cite{130}.

One of the most common classical mechanisms of spontaneous symmetry breaking
can be formalized in the following way.  The ground state is represented by a
vector $x_0$ belonging to the Euclidean space $\real^n$, on which $G$ acts as
a group of linear transformations; $x_0$ is determined as the point at which a
$G$-invariant potential $V(x)$ assumes its absolute minimum ($V$ might be a
Higgs potential in a gauge field theory or a thermodynamic potential in a
Landau theory of structural phase transitions), and $G_0$ is the isotropy
subgroup of $G$ at $x_0$ (the little group of $G_0$).  Generally, the
potential depends also on parameters $\gamma$ (for instance scalar
self-couplings in Higgs potentials, or pressure and temperature in
thermodynamic potentials), that cannot be determined from invariance
requirements under transformations of $G$.  In this case $x_0$ and $G_0$ can
depend on the $\gamma$'s, and various patterns of spontaneous symmetry
breaking are allowed, corresponding to distinct structural {\it phases} of the
system.

In supersymmetric field theories the absolute minimum of the potential
controls both the spontaneous symmetry and supersymmetry breaking (see, for
instance \cite{822} and references therein), and often the features of the two
breaking schemes are related \cite{220}.

In all the cases just mentioned, the determination of the ground state of the
system rests on a precise determination of the point $x_0$, where the
potential takes on its absolute minimum, and the determination has to be
analytical, since the isotropy subgroups of $G$ at nearby points may be
different.

Even if trivial in principle, the analytical determination of the minimum of
an invariant potential is generally a difficult computational task (even if
one uses polynomial approximations for the potential), owing to the large
number $n$ of the variables $x_i$ which are often involved.  An additional
difficulty is related to the degeneracy of the stationary points of the
potential, which is an unavoidable consequence of the invariance properties of
the potential; it prevents, in fact, a direct application \cite{555} of
Morse's theory \cite{600}.  Also the use of an extended Morse theory
\cite{070} seems not to give big advantages \cite{289}.

In 1971, Gufan \cite{278,277} proposed a new, more economical, approach to the
problem, which was based on the remark that a $G$-invariant function $V(x)$
can be expressed as a function $\widehat V(p_1, \dots ,p_q)$ of a finite set
$p(x)=(p_1(x),\dots ,p_q(x))$ of basic polynomial invariants.  When the point
$p\in\real^q$ ranges in the domain spanned by $p(x),\ x\in \real^n$, the
function $\widehat V(p)$ has the same range as $V(x)$, but is not plagued by
the same degeneracies.  Gufan's proposal found immediate applications in
crystal field theory (see refs.  \cite{333,450,453,454,455,542}, to cite but a
few of the pioneering papers on the subject).  A full and correct exploitation
of his idea required, however, an exact determination of the ranges of the
functions $p_i(x)$, a non trivial problem that was solved only ten years
later, when it was independently remarked \cite{020} that any $G$-invariant
function, being a constant along each orbit of $G$, can be considered a
function in the orbit space $\real^n/G$ of the action of $G$ in $\real^n$.  As
a consequence, the problem of determining the stationary points of $V(x)$
could be more economically reformulated in $\real^n/G$ \cite{021}, where the
$p_i$'s can be used advantageously to parametrize the orbits.  In $\real^n/G$,
the images of all the points of $\real^n$ with the same invariance properties
under $G$ transformations form smooth sub-manifolds, which are usually called
{\em strata}.  By varying the parameters $\gamma$, the location of the minimum
of $V(x;\gamma)$ may shift to a different stratum, thus causing a (structural)
phase transition of the system.

A sensible progress in the characterization of the geometry of the orbit
spaces of the CLG's was achieved using the powerful tools of geometric
invariant theory \cite{610,611}, which led to the discovery of a simple recipe
allowing to build a concrete image of the orbit space of any linear CLG and
its stratification \cite{020,021,680,651}.  It was shown that the orbit spaces
of the CLG's are connected semi-algebraic varieties, whose defining equations
and inequalities can be expressed in the form of positivity conditions of
matrices $\widehat P(p)$ built only in terms of the gradients of the basic
polynomial invariants $p_1(x), \dots ,p_q(x)$:

\eq{\widehat P_{ab}(p(x))\,=\,\summ 1ni \frac{\partial p_a(x)}{\partial x_i}
\frac {\partial p_b(x)}{\partial x_i}, \qquad a,b\,=\,1,\dots q.}

Using this result, one can obtain, for instance, a concrete realization of the
orbit space of any {\em coregular}\footnote{$G$ is said to be coregular if
there is no algebraic relation among the elements of a minimal set of its
basic invariant polynomials, i.e., among the elements of the minimal integrity
bases of the ring of its polynomial invariants.} finite linear group.  In
fact, the class of these groups has been shown to coincide with the class
formed by the finite groups generated by reflections (which are almost all
crystallographic groups) and explicit or algorithmic descriptions of their
basic polynomial invariants have been given by many authors (see for instance
\cite{134,135,173,174,304,543,670}).

For a general CLG, the matter is not that simple, since the determination of a
minimal complete set of basic invariant polynomials, i.e.\ of a minimal
integrity basis of the ring of polynomial invariants of $G$, may be a
difficult problem to solve.\footnote {A complete classification of compact
coregular linear groups is known, at present, only for finite groups and for
simple \cite{711} and semisimple \cite{458} Lie groups.} This serious handicap
in the direct approach to the determination of the $\widehat P$-matrix
associated to a general CLG stimulated the research, and led to the discovery,
of an alternative indirect method of computation of the $\widehat P$-matrices
associated to CLG's.  These matrices have been shown \cite{681,682,683} to be
solutions of a {\em master} differential equation, satisfying convenient
initial conditions ({\em allowable solutions}).  The master equation assumes a
particularly simple canonical form ({\em canonical equation}) for compact
coregular linear groups (hereafter abbreviated in CCLG's).  The form of the
canonical equation is the same for all CCLG's; it does involve only the
degrees of the elements of the integrity bases as free parameters.

The master equation approach to the determination and classification of the
$\widehat P$-matrices gives a strong support to the conjecture that the orbit
spaces of all the compact linear groups possessing a basis of $q$ independent
basic polynomial invariants with the same degrees can be classified in a
finite (and small, for small dimensions and degrees) number of isomorphism
classes.  The conjecture has been proved to hold true for $q\le 4$.

This fact makes the orbit space approach to the study of covariant functions,
and in particular of spontaneous symmetry breaking, particularly appealing.
In fact, invariance properties are often the only bounds which are imposed on
the potential (beyond regularity and stability properties and/or bounds on the
degree, when the potential is a polynomial function).  If the symmetry groups
of the potentials of different theories share isomorphic orbit spaces, the
potentials have the same formal expression and the same domain when written as
functions in orbit space, despite the completely different physical meaning of
the variables and parameters involved in the definition of the potentials.
Thus, the problems of determining the geometric features of the phase space,
the location and stability properties of the minima of the potential, the
number of primary strata (and, consequently, the maximum number of phases) and
the allowed transitions between primary strata are identical in all these
theories \cite{021,321,681,682}.

The pursuit of the ambitious program of determining the orbit spaces of all
the CLG's, following the master equation approach, has already given
encouraging results, but has left some serious open problems \cite{683}.  The
main ones are listed below; some of them will be dealt with and partially
solved in this paper:

\begin{enumerate}
\item
All the allowable solutions of the canonical equation have been determined for
$q\le 4$ (see refs.  \cite{682,685} hereafter referred to as I and II), while
for $q>4$, the determination of all the allowable solutions appears to be
still possible, but extremely lengthy.  The set-up of an inductive procedure
for the determination of at least a part of the allowable solutions of the
canonical equation is in progress \cite{684,686}.

\item
The canonical equation and the associated initial conditions are only a set of
necessary conditions that the $\widehat P$-matrices of the CLG's must satisfy;
even if quite stringent, they need not be sufficient.  Therefore, once the
allowable solutions of the canonical equation have been determined, the
problem remains of selecting those which are really generated by a group.  In
this paper we shall give a partial answer to this problem in the case of
coregular groups with $q\le 4$.

\item
An effective formalization of the condition that there is no algebraic
relation among the elements of a minimal integrity basis (minimality +
regularity condition) has not yet been found, nor used, in I and II.  Thus, it
cannot be excluded that some of the allowable $\widehat P$-matrices determined
in I and II are indeed associated to non minimal bases or to non-coregular
groups.

\item
A sound analysis of the structure of the master equation in the general
non-coregular case is still missing; some results have been obtained only for
non-coregular CLG's with a sole independent relation among the elements of the
minimal integrity bases \cite{687}.
\end{enumerate}

\vskip3truemm
The paper will be organized in the following way.  We shall begin, in \S 2,
with a short survey of the geometry of linear group actions, of the properties
of the canonical equation, and we shall briefly argue on the possibility of
classifying the orbit spaces of the CCLG's through the determination of the
allowable solutions of the canonical equation.  In \S 3 and \S 4 we shall
determine explicitly the $\widehat P$-matrices associated to all the finite
irreducible and, respectively, reducible reflection groups with no more than
four independent basic invariants.  The results will be obtained using the
explicit form of the basic invariant polynomials of the reflection groups that
can be found in the mathematical and physical literature.  A comparison of our
results with the allowable solutions of the canonical equation reported in I
and II will allow us to identify generating groups for all the irreducible,
and some of the reducible, allowable $\widehat P$-matrices.

After a few concluding remarks on our mathematical results, collected in \S 5,
in the last section we shall illustrate how they can be used in one of the
specific physical contexts mentioned in the introduction.  Our aim will be to
show that, despite the sophisticated mathematical tools that have been used to
achieve the results presented in the paper, their practical exploitation in
physical contexts only requires an elementary use of standard analysis,
geometry and group theory.  The physical problem we shall deal with in \S 6
has been studied by various authors in the past \cite{Kim}; our revisitation
will not lead to essentially new results, but for the fact that the explicit
knowledge of the algebraic relations defining the strata will allow us to
arrive at explicit analytic solutions in a particularly simple way.

\section{An overview of the geometry of linear group actions}

In this section, we shall first define most of our notations and recall,
without proofs, some results concerning invariant theory and the geometry of
orbit spaces of CLG's (see for instance \cite{080,710} and references
therein), then we shall introduce the first definitions and the basic tools
for our subsequent analysis.

For our purposes, it will not be restrictive to assume that $G$ is a matrix
subgroup of $O_{n}(\real)$\footnote{The stronger assumption $G\subseteq
SO_n(\real)$ introduced in I and II is due to a slip; in fact, the
unimodularity condition has never been used in these references.} acting
linearly in the Euclidean space $\real^{n}$.

\subsection{Orbits and strata}
We shall denote by $x= (x_1,\dots ,x_n)$ a point of $\real^ n$. The group $G$
acts in $\real^n$ in the following way:

\eq{x'_i\,=\,(g\cdot x)_i\,=\,\summ 1nj g_{ij}\, x_j\,,\qquad x\in\real^n,\
g\in G\,.}
The $G$-orbit $\Omega_{\overline{x}}$ through $\bar x\in\real^n$ and the {\em
isotropy subgroup} $G_{\overline{x}}$ of $G$ at $\overline{x}\in\real^n$ are
defined by the following relations:

\eq{\Omega_{\overline{x}} \,=\, \{ g \cdot \overline{x} \mid g \in G\},\qquad
G_{\overline{x}} \,=\, \{g \in G \mid g \cdot \overline{x} = \overline{x}\}
\,.}

The invariance of the Euclidean norm under orthogonal transformations assures
that the G--orbit through $\overline{x}$ is contained in the sphere of radius
$\overline{x}$, centered in the origin of $\real^n$, while the linearity of
the action of $G$ in $\real^n$ implies

\eq{G_{\overline{x}} \,=\, G_{\lambda \overline{x}}, \,\, \forall \lambda \in
\real _*\,. }
The isotropy subgroup of $G$ at the origin of $\real^n$ coincides with $G$.
The isotropy subgroups $G_{g \cdot \overline{x}}$ of $G$, at points lying on
the same orbit $\Omega_{\overline{x}}$ are conjugated subgroups in $G$:

\begin{equation}
G_{g \cdot \overline{x}} \,=\, g G_{\overline{x}} g^{-1}, \, \forall g \in
G\,. \label{iso_group2}
\end{equation}

The class of all the subgroups of $G$ conjugated to $G_{\overline{x}}$ in $G$
will be said to be the {\em orbit type of} $\Omega_{\overline{x}}$ or,
equivalently, of the points of $\Omega_{\overline{x}}$; it specifies the
symmetry properties of $\Omega_{\overline{x}}$ under transformations induced
by elements of $G$.

The set of all the points $x\in\real^n$ (or, equivalently, of all the orbits
of $G$) with the same orbit type form an {\it isotropy type stratum of the
action of $G$ in} $\real^ n$, hereafter called simply a {\it stratum of}
$\real^n$.  All the connected components of a stratum are smooth
iso-dimensional sub-manifolds of $\real^n$.

\subsection{The orbit space}
The {\em orbit space} of the action of $G$ in $\real^ {n}$ is defined as the
quotient space $\real^ n/G$ (obtained through the equivalence relation between
points belonging to the same orbit) endowed with the quotient topology and
differentiable structure.  We shall denote by $\pi$ the canonical projection
$\real^n\rightarrow\real^ {n}/G $.  Whole orbits of $G$ are mapped by $\pi$
into single points of $\real^{n}/G$.  The image through $\pi$ of a stratum of
$\real^{n}$ will be called an ({\it isotropy type}) {\it stratum of}
$\real^{n}/G$; all its connected components turn out to be smooth
iso-dimensional manifolds.

Almost all the points of $\real^n/G$ belong to a unique stratum $\Sigma_p$,
the {\it principal stratum}, which is a connected open dense subset of $\real^
n/G$.  The boundary $\overline{\Sigma_p}\backslash\Sigma_p$ of $\Sigma_p$ is
the union of disjoint {\it singular} strata.  All the strata lying on the
boundary $\overline{\Sigma}\backslash\Sigma$ of a stratum $\Sigma$ of
$\real^n/G$ are open in $\overline{\Sigma}\backslash\Sigma$.

The following partial ordering can be introduced in the set of all the orbit
types:  $[H]<[K]$ if $H$ is a subgroup of a subgroup of $G$ conjugated with
$K$.  The orbit type $[H]$ of a stratum $\Sigma$ is contained in the orbit
types $[H_b]$ of all the strata $\Sigma_b$ lying in its boundary; therefore,
more peripheral strata of $\real^n/G$ are formed by orbits with higher
symmetry under $G$ transformations.  The number of distinct orbit types of $G$
is finite and there is a unique minimum orbit type, the {\it principal orbit
type}, corresponding to the principal stratum; there is also a unique maximum
orbit type $[G]$, corresponding to the image through $\pi$ of all the points
of $\real^n$, which are invariant under $G$.

A faithful image of $\real^n/G$ can be obtained making use of a basic result
of the geometric approach to invariant theory in the following way.

A function $f(x)$ is said to be $G$-invariant if

\eq{f(g\cdot x)\,=\,f(x),\qquad\forall x\in \real^n,\ g\in G.}
The set of all real, $G$-invariant, polynomial functions of $x$ forms a ring
$\real [x]^G$, that admits a finite integrity basis \cite{286,617}.
Therefore, there exists a finite minimal collection of invariant polynomials
$p(x) = ( p_1(x), p_{2}(x), \ldots, p_{q}(x) )$ such that any element
$F\in\real[x]^G$ can be expressed as a polynomial function $\widehat F$ of
$p(x)$:

\eqll{\widehat{F}(p(x)) \,=\, F(x), \, \forall x \in \real^{n}\,.}{e2}
The polynomial $\widehat F(p)$ will be said to have weight $w$, if $w$ is the
degree of the polynomial $F(x)$; it will be said to be $w$-homogeneous if such
is $F(x)$.

The elements of a (minimal) basis of $\real[x]^G$ can be chosen to be
homogeneous polynomials.  The number $q$ of elements of a minimal integrity
basis and their homogeneity {\em degrees} $d_i$'s are only determined by the
group $G$.

To avoid trivial situations, in this paper we shall only consider linear
groups with no fixed points, but for the origin of $\real^{n}$.  In this case,
the minimum degree of the elements of a minimal integrity basis is necessarily
2, and the following conventions can be adopted:

\begin{equation}
d_{1} \geq d_{2} \geq \ldots d_{q} \,=\, 2\,; \qquad
p_{q}(x) \,=\, \| x \|^2
 \,=\, \summ 1ni x_{i}^{2}\,.
\label{conventions}
\end{equation}
Hereafter, by a minimal integrity basis of $G$ (abbreviated into MIB of $G$)
we shall always mean a {\it minimal homogeneous integrity basis} for the ring
of $G$-invariant polynomials, for which the conventions of \eref{conventions}
hold true.

Since $G$ is a compact group, the orbits of $G$ are separated by the elements
of a MIB of $G$, i.e., at least one element of a MIB of $G$ takes on different
values on two distinct orbits.  Thus, the elements of a MIB of $G$ provide a
good parametrization of the points of $\real^{n}/G$, that turns out to be also
smooth, since the orbit map $p:\, \real^{n} \longrightarrow \real^{q}$, which
maps all the points of $\real^{n}$ lying on an orbit of $G$ onto a single
point of $\real^{q}$, induces a diffeomorphism of $\real^{n}/G$ onto a
semialgebraic q--dimensional connected closed subset of $\real^{q}$.

\subsection{Coregular and non--coregular groups}
The group $G$ is said to be {\em coregular} if the elements of its MIB's are
algebraically, and therefore functionally, independent.  If $G$ is
non-coregular, the elements of any one of its MIB's satisfy a certain number
of algebraic identities in $\real^n$:

\eqll{\widehat F_A(p(x))\,=\,0,\qquad A=1,\dots, K\,.}{e4}
The associated equations

\eqll{ \widehat F_A(p)\,=\,0,\qquad A=1,\dots, K,}{e5}
define an algebraic variety in $\real^q$, which will be called the {\em
variety $\cal Z$ of the relations} (among the elements of the MIB). The
number $K$ will be said the {\it coregularity order} of $G$. If $G$
is coregular, there are no relations among the elements of its MIB's, the
coregularity order of $G$ is zero and we shall set ${\cal Z}=\real^n$.

{}From now on, in this paper, we shall deal mainly with coregular CLG's.

\subsection{The $\widehat{P}(p)$ matrix}
A characterization of the image $p(\real^n)$ of the orbit space of $G$ as a
semi-algebraic variety can be easily obtained through a matrix
$\widehat{P}(p)$, defined only in terms of the $G$-invariant Euclidean scalar
products between the gradients of the elements of the MIB $\{p(x)\}$:

\begin{equation}
P_{ab}(x) \,=\, \summ 1ni\frac{\partial p_a(x)}{\partial x_i}\cdot\frac{
\partial p_b(x)}{\partial x_i} \,=\, \widehat{P}_{ab}(p(x)) \hspace{3em}
a,b = 1, \ldots, q\,; \label{matP}
\end{equation}
where in the last member, use has been made of Hilbert's theorem, in order to
express $P_{ab}(x)$ as a polynomial function of $p_1(x), \dots ,p_q(x)$.

The following fundamental theorem clarifies the meaning and points out the
role of the matrix $\widehat{P}(p)$:

\begin{theo}     \label{T1}
Let $G$ a compact coregular subgroup of $\mbox{O}_{n}(I\!\!R)$, $p$ the map
$\real^n\rightarrow\real^q$ defined by the homogeneous MIB
$\{p_1(x),p_2(x),\ldots ,p_q(x)\}$ and $\widehat{P}(p)$ the matrix defined in
\eref{matP}.  Then $\overline{{\cal S}}= p(\real^{n})$ is the unique
semialgebraic connected subset of the variety ${\cal Z}\subseteq \real^q$ of
the relations among the elements of the MIB where $\widehat{P}(p)$ is positive
semi-definite.  The $k$--dimensional primary strata of $\overline{{\cal S}}$
are the connected components of the set $\widehat{W}^{(k)} = \{ p \in {\cal
Z}; \mid \widehat{P}(p) \geq 0, \, \mbox{rank} (\widehat{P}(p)) = k \}$; they
are the images of the connected components of the k--dimensional isotropy type
strata of $\real^{n}/G$.  In particular the variety ${\cal S}$ of the interior
points of $\overline{{\cal S}}$, where $\widehat{P}(p)$ has the maximum rank,
is the image of the principal stratum.  \end{theo}

It will be worthwhile to note that, for coregular groups, ${\cal Z}=\real^q$
and that the image of the unit sphere of $\real^n$ under the orbit map $p(x)$
is a {\em compact connected} $(q-1)$-dimensional semialgebraic variety in
the space $\real^{q-1}$ spanned by the variables $p_1,\dots ,p_{q-1}$.

The following properties, which are common to all the matrices $\widehat
P(p)$, are more or less immediate consequences of the definition of these
matrices:

\begin{description}
\item[P1 {\em Symmetry, homogeneity and bounds on the last row and column:}]
\ The matrix \newline $\widehat{P}(p)$ is a $q \times q$ symmetric matrix,
whose elements $\widehat{P}_{ab}(p)$ are real w--homogene\-ous polynomials
with weight

\eqll{w(\widehat{P}_{ab}) \,=\, d_{a} + d_{b} - 2\,.}{weightP}
The last row and column are determined by the degrees of the MIB:

\eqll{ \widehat{P}_{qa}(p) \,=\, \widehat{P}_{aq}(p) \,=\, 2 d_{a} p_{a},
\qquad a=1, 2, \ldots, q\,.}{LastRow}

\item[P2 {\em Tensor character:}] The matrix elements of $\widehat{P}(p)$
transform as the components of a rank 2 contravariant tensor under MIB
transformations that maintain the conventions fixed in \eref{conventions}
(these transformations will be hereafter called MIBT's).  In fact, let $\{
p(x)\}$ and $\{p'(x)\}$ be distinct MIB's; the $p'_a(x)$'s, being G--invariant
polynomials, can be expressed as polynomial functions of the $p_a(x)$'s:

\begin{equation} \label{MIBT}
{p'}_{\alpha}  \,=\,  {p'}_{\alpha}(p) \qquad\alpha = 1, 2, \ldots, q - 1\,,
\end{equation}
where the polynomial function $p'_\alpha(p)$ only depends on the $p_a$'s such
that\footnote{Since in our conventions the $q$-th element
of any MIB is fixed to equal $\summ 1ni x_i^2$, when defining a MIBT we shall
always understand the condition $p'_q(p)=p_q$.} $d_a\le d'_\alpha $. Then,

\begin{equation} \label{equiv}
\widehat{P}'(p'(p)) \,=\, J(p) \cdot \widehat{P}(p) \cdot J^{T}(p) \,,
\end{equation}
where we have denoted by $J(p)$ the Jacobian matrix of the transformation:

\eq{J_{ab}(p) \,=\, \partial {p'}_{a}(p)/ \partial p_{b},\qquad a,b=1,\dots
,q\,;}
the matrix $J$ turns out to be upper--block triangular and the determinant of
$\widehat{P}(p)$ to be a relative invariant of the group of the MIBT's.
\end{description}

\subsection{Classification of the orbit spaces of CCLG's}
Two matrices $\widehat{P}(p)$ and $\widehat P'(p')$ will be said to be {\em
equivalent} if they are connected by a relation like \eref{equiv}, where
$J(p)$ is the Jacobian matrix of a MIBT $p'= p'(p)$.  Thus, the
$\widehat
P$-matrices computed from different MIB's of the same CLG are equivalent, and
the semialgebraic varieties $\overline{{\cal S}}$ and $\overline{{\cal S'}}$
defined by the positivity conditions imposed on $\widehat{P}(p)$ and
$\widehat P'(p')$ respectively, are equivalent concrete realizations of the
orbit space $\real^{n}/G$.

Since $G$ is coregular, its orbit space is completely determined by the
positivity conditions of a $\widehat P$-matrix computed from any one of its
MIB's; for non-coregular groups, also a complete set of relations among the
$p_a$'s has to be specified.

\subsection{Isomorphism classes of orbit spaces}
The notions of MIBT's (see \eref{MIBT}) and of equivalence of
$\widehat{P}$-matrices (see \eref{equiv}) can be extended to the case of
different coregular groups $G$ and $G'$, provided that their MIB's have the
same number of elements, with the same degrees. Let, $\{p\}$ and $\{p'\}$ be
MIB's for $G$ and $G'$, respectively.

\begin{defin}
The orbit space $\real^{n}/G$ and $\real^{n'}/G'$ of the compact coregular
linear groups $G$ and $G'$ will be said to be isomorphic if the associated
$\widehat P$-matrices $\widehat{P}(p)$ and $ \widehat P'(p')$ satisfy
\eref{equiv}, where the transformation $p'= p'(p)$ has all the formal
properties of a MIBT.  \end{defin}

If $G$ and $G'$ have isomorphic orbit spaces, then the images of their orbit
spaces $\overline{{\cal S}}$ and $\overline{{\cal S'}}$, associated with the
MIB's $\{p\}$ and $\{p'\}$ are isomorphic semialgebraic varieties:

\eq{\overline{{\cal S'}} \,=\, p'(\overline{{\cal S}})\,.}

Thus the classification of the isomorphism classes of the orbit spaces of the
CCLG's rests on the determination of a representative for each class of
equivalent $\widehat{P}(p)$ matrices (and, for non--coregular groups, on the
determination of the possible relations among the elements of the MIB's).
This can be done, in principle, for all CCLG's.  The matrices $\widehat{P}(p)$
have been shown \cite{681}, in fact, to be solutions of a {\em canonical
differential equation}, satisfying convenient initial conditions ({\em
allowable} solutions).  The canonical equation does involve only the degrees
$\{ d_{1}, d_{2}, \ldots, d_{q} \}$ of the MIB's as free parameters, as we
shall see in the next subsection.

\subsection{Boundary and positivity conditions}
The orbit space $\overline{{\cal S}}$, defined in Theorem~\eref{T1}, is a
connected $q$-dimensional semialgebraic variety of $\real^{q}$ and, like all
semialgebraic varieties \cite{840}, it presents a natural stratification in
connected semialgebraic sub-varieties $\hat{\sigma}$, called {\em primary
strata}\footnote{A simple example of a compact connected semialgebraic
variety of $\real^3$ is yielded by a polyhedron.  Its interior points form its
unique 3-dimensional primary stratum, while 2-,1- and 0-dimensional primary
strata are formed, respectively, by the interior points of each face, by the
interior points of each edge, by each vertex.} We shall denote by ${\cal
I}(\hat{\sigma})$ the ideal formed by all the polynomials in $p \in \real^{q}$
vanishing on $\hat{\sigma}$.  Every $\hat{f}(p)\in {\cal I}(\hat{\sigma})$
defines an invariant polynomial function in $\real^{n}$ vanishing at all
the points $x$ lying in the set $\Sigma_{f}=p^{ -1}(\hat{\sigma})$:

\begin{equation}
f(x) \,=\, \hat{f}(p(x)) \,=\, 0, \hspace{4em} \forall x \in \Sigma_{f}\,.
\end{equation}

The gradient $\partial f(x)$ is obviously orthogonal to $\Sigma_f$ at
every $x\in\Sigma_f$, but, it must also be tangent to $\Sigma_{f}$
since $f(x)$ is a G--invariant function \cite{080,680}. As a consequence,

\begin{equation} \label{A1}
0\,=\,\partial f(x) \,=\, \left.\summ 1qb\partial_b \hat{f}(p)\right|_{p=p(x)}
\cdot \partial p_{b}(x)\,, \hspace{2em} \forall x \in \Sigma_{f} \,.
\end{equation}
By taking the scalar product of \eref{A1} with $\partial p_{a}(x)$, we
end up with the following {\em boundary conditions} \cite{683}:

\begin{equation} \label{A2}
\summ 1qb \widehat{P}_{ab}(p) \, \partial_b \hat{f}(p) \in
{\cal I}(\hat{\sigma}), \hspace{2em} \forall \hat{f} \in {\cal I}
(\hat{\sigma}) \,\, \mbox{and}
\,\, \forall \hat{\sigma} \subseteq \overline{{\cal S}}\,.
\end{equation}
If $\{ Q_{1}^{(\hat{\sigma})}(p), Q_{2}^{(\hat{\sigma})}(p), \ldots,
Q_{m}^{(\hat{\sigma})}(p) \}$ is an integrity basis for ${\cal
I}(\hat{\sigma})$, \eref{A2} is equivalent to:

\begin{equation} \label{A3}
\summ 1qb \widehat{P}_{ab}(p) \, \partial_b Q_{r}^{(\hat{\sigma})}(p)
 \,=\, \summ 1ms \lambda_{rs;a}^{(\hat{\sigma})}(p) \,
Q_{s}^{(\hat{\sigma})}\,,
\end{equation}
where the $\lambda$'s are $w$-homogeneous polynomial functions of $p$.

In the particular case in which $\hat\sigma$ is a $(q-1)$-dimensional primary
stratum, the ideal ${\cal I}(\hat\sigma)$ has a unique {\it irreducible}
generator, $Q^{(\hat\sigma)}(p)$, and \eref{A3} reduces to the simpler form
\cite{681,683}

\begin{equation} \label{2.7.1}
\summ 1qb \widehat P_{ab}(p) \, \partial_b Q^{(\hat{\sigma})}(p)
 \,=\, \lambda^{(\hat{\sigma})}_a(p)\, Q^{(\hat{\sigma})}(p)\,,\qquad a=1,
\dots, q\,.
\end{equation}

The validity of \eref{A3} can be extended to the case in which $\hat\sigma$ is
a union of primary strata.  In particular, the ideal ${\cal I} ({\cal B})$,
associated to the union $\cal B$ of all the $(q-1)$-dimensional strata of
$\overline{\cal S}$ (whose closure forms the boundary of $\overline{\cal S}$)
has a unique generator $A(p)$:

\eqll{A(p)\,=\,\prod_{\hat\sigma\subseteq{\cal B}}Q^{(\hat\sigma)}(p)}{2.7.3}
and the following relation is satisfied:

\eqll{\summ 1qb \widehat{P}_{ab}(p) \, \partial_b A(p) \,=\,
\lambda^{(A)}_a(p) \, A(p) \,,}{A4}
where $\lambda^{(A)}(p)$ is a contravariant vector field with $w$-homogeneous
components, and

\eq{\lambda^{(A)}(p)\,=\,\sum_{\hat{\sigma}\subseteq{\cal
B}}\lambda^{(\hat{\sigma})}(p)\,.}

The results summarized below have been proved in \cite{682}.

The vector $\lambda^{(A)}(p)$ can be reduced to the canonical form
$\lambda_a^{(A)}(p)=2\delta _{aq}w(A)$ in particular MIB's, the so-called {\it
$A$-bases}, which are intrinsically defined.  In an $A$-basis, the boundary
conditions assume the following {\it canonical} form:

\eqll{\summ 1qb \widehat P_{ab}(p)\partial_b A(p) \,=\, 2\delta
_{aq}w(A)A(p),\qquad a=1,\dots ,q\,.}{a1}

{}From \eref{a1} one deduces that, in every $A$-basis, the following
facts hold true:

\vskip3truemm\begin{description}
\item[i)] The point $p^{(0)}\,=\,(0,\dots , 0,1)$ lies in $\cal S$; it is the
image of a $G$-orbit lying on the unit sphere of $\real^n$.
\item[ii)] $A(p)$ is a factor of $\det \widehat P(p)$; its weight is bounded:

\eqll{2d_1\le w(A)\le w(\det\widehat P)\,=\,2\summ 1qa d_a -2q}{a2a}
and it can be normalized at $p^{(0)}$:

\eqll{A(p^{(0)})\,=\,1\,;}{a2b}
we shall call it the {\it complete active factor} of det$\widehat P(p)$.

\item[iii)] The restriction, $A(p)\big |_{p_q=1}$, of $A(p)$ to the image of
the unit sphere of $\real^n$ in $\real^q$ has a unique local non degenerate
maximum lying at $p^{(0)}$; thus:

\eqll{\left .\partial_\alpha A(p)\right|_{p=p^{(0)}}\,=\,0,\qquad \alpha =1,
\dots, q-1\,.}{gradA}

\item[iv)] $\widehat P(p^{(0)})$ is block diagonal, each block being
associated to a subset of $p_a$'s sharing the same degree, and, in a subclass
of $A$-bases ({\it standard $A$-bases}), it is diagonal:

\eqll{\widehat P_{ab}(p^{(0)})\,=\,d_ad_b\delta_{ab}, \qquad a,b=1, \dots,
q.}{iniP}
Two different standard $A$-bases are related by a MIBT not involving $p_q$:

\eqll{p'_\alpha \,=\,f_\alpha (p_1,\dots ,p_{q-1}),\qquad \alpha = 1, \dots ,
q-1\,;}{a2d}
the corresponding Jacobian matrix is orthogonal at $p^{(0)}$.
\end{description}

\subsection{The canonical equation}

\vskip3mm Let us now look at the boundary conditions from a different point of
view:  the set $\{p_1,\dots ,p_q\}$ will be viewed as a set of {\em weighted
indeterminates}, with integer weights $d_1, \dots ,d_q$ satisfying the
conventions

\eqll{d_{1} \geq d_{2} \geq \ldots d_{q} \,=\, 2\, ;}{a3a}
and the boundary conditions expressed in \eref{a1} will be considered as a set
of equations in which $\widehat P_{ab}(p)$ and $A(p)$ are thought of as
unknown polynomial functions of $p$, satisfying the conditions listed under
items {\bf P1-P2} and in \eref{a2a}.  The positivity conditions specified in
\eref{a2b} and \eref{iniP} will be treated as initial conditions.  With the
above meaning for the symbols, {\eref{a1} will be called the {\it canonical
equation}.

The solutions $(\widehat P(p), A(p))$ of the canonical equation satisfying the
initial conditions specified in \eref{a2b} and \eref{iniP} will be called {\it
allowable solutions} and the corresponding $\widehat P$ matrices, {\em
allowable $\widehat P$ matrices}.

A solution of the canonical equation will be said to be {\it irreducible}, if
$A(p)$ is an irreducible (real) polynomial and {\it fully active}, if
$A(p)={\rm const}\cdot {\rm det}\widehat P(p)$.

Two allowable $\widehat P$ matrices will be said to be equivalent if a
$w$-homogeneous transformation exists on the indeterminates
$p_1,\dots,p_{q-1}$ such that \eref{equiv} is satisfied.

The allowable $\widehat P$-matrices $\left.  \widehat P(p)\right|_{p_q=1}$
have been shown to be positive semi-definite only in a compact
$(q-1)$-dimensional semialgebraic variety of the space $\real^{q-1}$ spanned
by the variables $p_1, \dots ,p_{q-1}$, containing the point $p^{(0)}$.

All the $\widehat P$-matrices associated to the different MIB's of any CCLG
with no fixed points are necessarily equivalent to an allowable $\widehat
P$-matrix.  At present we do not know if the converse holds also true, i.e.,
if every {\em allowable} $\widehat P$-matrix is generated by a CCLG with no
fixed points.

The allowable solutions of the canonical equations for $q \leq 4$ have been
determined in I and II.  For each choice of the degrees $\{ d_{1}, d_{2},
\ldots, d_{q} \}$, only a finite or null number of non equivalent solutions
has been found, showing the existence of {\em selection rules} for the degrees
of the CCLG's. The solutions can be organized in {\em towers}; the degrees of
the elements of the same tower can be written in the form $d_\alpha=sd_\alpha^
{(0)},\ \alpha= 1,\dots ,q-1,$ where $s$ is an integer scale parameter. A
solution of the canonical equation corresponding to $s=1$ will be said a {\it
fundamental} solution.

\vskip3mm
Exploiting the fact that all the {\em finite} coregular linear groups have
been classified in the mathematical literature and the associated MIB's have
been determined \cite{731}, in the following sections we shall determine the
$\widehat P$-matrices of all the finite CLG's with 2-, 3- \cite{683} and
4-dimensional \cite{685} orbit spaces, and we shall check that they can all be
found among the allowable solutions of the canonical equation listed in I and
II.

\section{Irreducible reflection groups with 2, 3 and 4 basic invariants}

Finite groups generated by reflections exhaust the class of {\em finite}
CLG's.  The explicit form of the elements of at least one MIB is known for all
these groups \cite{135,173,174,304,543,670}.  Thus, the corresponding
$\widehat{P}(p)$ matrices can be computed, as well as the complete factors
$A(p)$ of $\det \widehat{P}(p)$ and the vector fields $\lambda^{(A)}(p)$
appearing in \eref{A4}.

In general, the MIB's proposed in the literature do not correspond to
$A$--bases, so the comparison with the results reported in I and II is not
immediate.  The easiest way to determine the form of a MIBT leading to an
$A$--basis is through the following condition on the Jacobian matrix
$J_{ab}(p)$ of the transformation:

\begin{equation} \label{SP1}
{\lambda'}^{(A)}_{a}(p'(p)) \,=\, \summ 1qb J_{ab}(p)\, \lambda^{(A)}_{b}(p)
\,=\, 0,\qquad a = 1, 2, \ldots, (q-1)\,.
\end{equation}

When the $A$--basis is not unique, \eref{SP1} is not sufficient to determine
all the free parameters involved in the definition of the MIBT.  The residual
free parameters can however be determined by requiring that the parametric
expression of $\widehat P'(p')$ in a general A--basis coincides with an
allowable $\widehat{P}$-matrix listed in I or II.  To shorten our formulas and
to make easier the comparison, we shall define

\eqll{\tilde p\,=\,(p_1,\dots ,p_{q-1}, 1)\,,\qquad {\rm for}\ p\,=\,(p_1,
\dots ,p_q)\,,}{p-tilda}

\eqll{\check x\,=\,(x_1,\dots ,x_{n}, 0)\,,\qquad {\rm for}\ x\,=\,(x_1,\dots
,x_n)\,,}{x-check}

\eqll{R_{ab}\,=\,\widehat P_{ab}/d_ad_b, \qquad a,b=1,\dots ,q,}{R}

\eqll{f_{n,k}(x)\,=\,\summ 1ni x_i^{n+1-k}\,,\qquad {\rm for}\
x\,=\,(x_1,\dots ,x_n)\,.}{f}

\subsection{Irreducible CCLG's with 2 basic invariants}

The irreducible CCLG's with 2-dimensional orbit spaces are classified in the
mathematical literature according to the following {\em types}:  A$_2$, B$_2$,
G$_2$ and I$_2(m)$.  The type G$_2$ and I$_2(6)$ groups have the same
invariants, so we shall not discuss separately the group of type G$_2$.

\vskip3truemm

\subsection{Type A$_2$}
The group acts on the plane $y_1+y_2+y_3=0$ of $\real^3$ by permutations.
Therefore the group and its invariants can be obtained from the reduction
of the group S$_{3}$, acting on $y=(y_1, y_2, y_3)$ by permutations of the
coordinates.

A MIB for S$_3$ is yielded by $\{f_{3,1}(y), f_{3,2}(y), f_{3,3}(y)\}$, where
$f_{3,k}$ is defined in \eref{f}.  The reduction of the linear group $S_3$ can
be obtained by means of an orthogonal basis transformation in $\real^{3}$,
induced by the matrix

\begin{equation}
A \,=\, \frac{1}{6} \left ( \begin{array}{cccc}
			\sqrt{3} & - \sqrt{3} & 0  \\ [1em]
			     1   &  1         & -2  \\ [1em]
		       \sqrt{2}  & \sqrt{2} & \sqrt{2}
\end{array}  \right )\,.
\end{equation}
The elements of the linear group of type A$_2$ are obtained as the principal
minors built with the first 2 rows and columns of the matrices $AgA^{-1},\
g\in S_3$ and, after setting, for $x=(x_1, \dots ,x_n)$,

\eqll{p_a(x)\,=\,\tilde p_a(A^{-1}\check x)\,,}{pa}
and $x=(x_1,x_2)$, a MIB for A$_2$ is yielded by $\{p_1(x), p_2(x)\}$ or,
explicitly:

\begin{eqnarray}
p_{1}(x) &=&  x_2 \left (3 x_1^2 - x_2^2\right)/ \sqrt{6}\,,\nonumber\\
p_{2}(x) &=&  x_{1}^{2} + x_{2}^{2} \,.
\end{eqnarray}

The unique element of the associated reduced $\widehat P$-matrix is easily
calculated to be

\eq{ \widehat P_{11}(p) \,=\, 3p_2^2/2 }
and after the following MIBT:

\eq{p'_1\,=\,\sqrt{6}\,p_1\,,}
one obtains, for $p'_2=1$, the allowable $\widehat{P}(p)$ matrix determined in
I for $q=2$ and $d_1=3$:

\eq{R'_{11}(\tilde p') \,=\, 1\,.}

\subsection{Type B$_2$}
The group acts on $x=(x_1, x_2)$ by permutations and sign changes of the
coordinates.  A MIB can be chosen as follows:

\begin{eqnarray}
p_{1}(x) &=& x_{1}^{4} + x_{2}^{4}\,, \nonumber \\
p_{2}(x) &=& x_{1}^{2} + x_{2}^{2} \,.
\end{eqnarray}
The unique essential element of the associated $\widehat P$-matrix turns out
to be the following

\eq{ \widehat{P}_{11}(p) \,=\, 8p_2\left(3p_1- p_2^2\right) }
and after the following MIBT:

\eq{p'_1\,=\, 4p_1-3p_2^2\,,}
one obtains, for $p'_2=1$, the allowable $\widehat{R}(p)$ matrix determined in
I for $q=2$ and $d_1=4$:

\eq{R'_{11}(\tilde p') \,=\, 1\,.}

\subsection{Type I$_2(m)$,  $m\ge 3$}
The groups of type I$_2(m)$, $m\ge 3$, are the dihedral groups ${\cal D}_m$,
defined as the groups of orthogonal transformations of $\real^2$ which
preserve a regular $m$--sided polygon centered at the origin.  After setting

\eq{z\,=\,x_1+ix_2\,,}
a MIB of ${\cal D}_m$ is yielded, for instance, by the invariants
$p(x)=\{p_1(x), p_2(x)\}$, with

\eq{p_1(x)\,=\,\Re{z^m},\qquad p_2(x)\,=\,|z|^2\,=\,x_1^2+x_2^2\,.}

The unique essential element of the associated $\widehat P$-matrix turns out
to be the following:

\eq{\widehat P_{11}(p)\,=\,d_1^2p_2^{m-1}.}
A comparison with I shows that $\{p\}$ is a standard A-basis.

\subsection{Type A$_3$}
The group acts on the plane $y_1+y_2+y_3+y_4=0$ of $\real^4$ by permutations.
Therefore the group and its invariants can be obtained from the reduction
of the group S$_4$, acting on $y=(y_1, y_2, y_3,y_4)$ by permutations of the
coordinates.

A MIB for S$_4$ is yielded by $\{f_1(y), \dots , f_4(y)\}$, where $f_k$ is
defined in \eref{f}.  The reduction of the linear group $S_4$ can be obtained
by means of an orthogonal basis transformation in $\real^{4}$, induced by the
matrix

\begin{equation}
A \,=\, \frac{1}{6} \left ( \begin{array}{cccc}
		       3 \sqrt{2} & -3 \sqrt{2} & 0 & 0 \\ [1em]
		       \sqrt{6} &  \sqrt{6}  & -2 \sqrt{6} & 0 \\ [1em]
		       \sqrt{3}  & \sqrt{3} & \sqrt{3} & - 3 \sqrt{3} \\ [1em]
			  3  &   3   &   3 &  3
\end{array}  \right )\,.
\end{equation}
The elements of the group of type A$_3$ are obtained as the principal minors
built with the first 3 rows and columns of the matrices $AgA^{-1},\ g\in S_4$
and, after setting $x=(x_1,x_2,x_3)$, a MIB is yielded by $\{p_1(x), p_2(x),
p_3(x)\}$, with $p_a(x)$ defined in \eref{pa}; explicitely:

\begin{eqnarray} \label{NA1}
 p_{1}(x) & = & \left ( 6 x_{1}^{4} + 12 x_{1}^{2} x_{2}^2 + 6 x_{2}^4
      + 12 \sqrt{2}\, x_{1}^{2} x_{2} x_{3} - 4 \sqrt{2}\, x_{2}^{3} x_{3}
      + 6 x_{1}^{2} x_{3}^{2} + 6 x_{2}^{2} x_{3}^{2} + 7 x_{3}^{4}
      \right)/12\,,\nonumber\\
 p_{2}(x) & =&  \left(3 \sqrt{2}\, x_{1}^{2} x_{2} - \sqrt{2}\, x_{2}^{3}
      + 3 x_{1}^{2} x_{3} + 3 x_{2}^{2} x_{3} - 2 x_{3}^{3} \right)/2\sqrt{3}
     \,,\nonumber  \\
 p_{3}(x) & = & x_{1}^{2} + x_{2}^{2} + x_{3}^{2} \,.
\end{eqnarray}
The essential elements of the associated $\widehat{P}(p)$-matrix turn out to
be the following:

\begin{eqnarray}
\widehat{P}_{11}(p) &=& 2\,\left(18\, p_{1} p_{3} + 2\, p_{2}^{2} -
		       3\, p_{3}^{3}\right)/3\,,\nonumber\\
\widehat{P}_{12}(p) &=& 7 p_{2} p_{3}\,,\nonumber\\
\widehat{P}_{22}(p) &=& 9\,\left(4\, p_{1} - p_{3}^{2}\right)/4
\end{eqnarray}
and after the following MIBT:

\begin{eqnarray}
p'_{1} &=& 12 p_{1} - 5 p_{3}^{2}\,, \nonumber \\
p'_{2} &=& 2 \sqrt{3}\, p_{2}\,,
\end{eqnarray}
one obtains, for $p'_3=1$, the matrix $R$ of class III.1($m = 1$), reported in
I:

\begin{eqnarray}
R'_{11}(\tilde p') &=& -{p'}_1 + {p'}_2^2 + 2\,,\nonumber\\
R'_{12}(\tilde p') &=& 2 {p'}_2  \,,\nonumber\\
R'_{22}(\tilde p') &=& p'_{1} + 2\,.
\end{eqnarray}

\subsection{Type B$_3$}
The group acts on $x=(x_1, x_2, x_3)$ by permutations and sign changes of the
coordinates.  A MIB can be chosen as follows:

\begin{eqnarray}\label{B3}
p_{1}(x) &=& x_{1}^{6} + x_{2}^{6} + x_{3}^{6}  \,,\nonumber \\
p_{2}(x) &=& x_{1}^{4} + x_{2}^{4} + x_{3}^{4} \,,\nonumber \\
p_{3}(x) &=& x_{1}^{2} + x_{2}^{2} + x_{3}^{2} \,\,.
\end{eqnarray}
The  essential elements of the associated $\widehat{P}(p)$-matrix turn out to
be the following:

\begin{eqnarray}\label{PB3}
\widehat{P}_{11}(p) &=& 30 p_{1} p_{2} + 30 p_{1} p_{3}^{2} -
30 p_{2} p_{3}^{3} + 6 p_{3}^{5}  \,,\nonumber \\
\widehat{P}_{12}(p) &=& 32 p_{1} p_{3} + 12 p_{2}^{2} - 24 p_{2}
p_{3}^{2} + 4 p_{3}^{4} \,,\nonumber \\
\widehat{P}_{22}(p) &=& 16 p_{1} \,\,.
\end{eqnarray}
and after the following MIBT:

\begin{eqnarray}
p'_{1} &=& 324 p_{1} - 432 p_{2} p_{3} + 124 p_{3}^{3} \,,\nonumber \\
p'_{2} &=& 18 p_{2} -10 p_{3}^{2} \,,
\end{eqnarray}
one obtains, for $p'_3=1$, the matrix $R$ of class III.2($m=1$) reported in I:

\begin{equation}
\begin{array}{rcl}
R'_{11}(\tilde p') &=& -{p'}_1 {p'}_2 - 4 {p'}_1 + 8 {p'}_2^2 - 16 {p'}_2
+ 64 \,,\nonumber \\
R'_{12}(\tilde p') &=& - 2 {p'}_1+{p'}_2^2+ 12 {p'}_2 \,, \nonumber \\
R'_{22}(\tilde p') &=& {p'}_1 + 4 {p'}_2  + 16 \,. \\
\end{array}
\end{equation}

\subsection{Type D$_3$}
The group acts on $x=(x_1, x_2, x_3)$ by permutations and by changes of an
even number of signs of the coordinates.  A MIB can be chosen as follows:

\begin{eqnarray}
p_1 &=& x_1^{4} + x_2^{4} + x_3^{4} \,,\nonumber \\
p_2 &=& x_1 \, x_2 \, x_3 \,,\nonumber \\
p_3 &=& x_1^2 + x_2^2 + x_3^2 \,.
\end{eqnarray}
The essential elements of the associated $\widehat P$-matrix turn out to be
the following:

\begin{eqnarray}
\widehat{P}_{11}(p) &=& 24 p_{1} p_{3} + 48 p_{2}^{2} - 8 p_{3}^{3} \,,
\nonumber\\
\widehat{P}_{12}(p) &=&  4 p_{2} p_{3} \,,\nonumber\\
\widehat{P}_{22}(p) &=&  (- p_{1} + p_{3}^{2})/2
\end{eqnarray}
and, after the following MIB transformation:

\begin{eqnarray}
p'_{1} &=& 4 p_{3}^{2} - 6 p_{1}  \,,\nonumber \\
p'_{2} &=& 6 \sqrt{3}\, p_{2} \,,
\end{eqnarray}
one obtains, for $p'_3=1$, the matrix $R$ of class III.1($m =1$) reported in
I.  The orbit spaces of the linear groups A$_3$ and D$_3$ turn out to be
isomorphic.

\subsection{Type H$_3$}
The group is the symmetry group of the icosahedron in $\real^3$.

Let us denote by $\tau$ the golden ratio:
\eqll{\tau\,=\,(1 + \sqrt{5})/2.}{tau}
Then, according to \cite{543}, a MIB for the group can be chosen as follows:
\begin{eqnarray}
 p_{1}(x) & = & (1+\tau^2)^{-5}\left[ (1 + \tau^{10})(x_{1}^{10} + x_{2}^{10}
+   x_{3}^{10}) + 45\, \tau^{2} ( x_{1}^{2} x_{2}^{8} + x_{2}^{2} x_{3}^{8} +
  x_{3}^{2} x_{1}^{8})  \right.\nonumber\\
 & & \hspace{2em} +210\, \tau^{4} ( x_{1}^{4} x_{2}^{6} +  x_{2}^{4} x_{3}^{6}
+ x_{3}^{4}
    x_{1}^{6} ) + 210\, \tau^{6} ( x_{1}^{6} x_{2}^{4} + x_{2}^{6} x_{3}^{4} +
    x_{3}^{6} x_{1}^{4})\nonumber\\
 & & \hspace{2em} \left. + 45\, \tau^{8}( x_{1}^{8} x_{2}^{2} + x_{2}^{8}
x_{3}^{2} +
   x_{3}^{8} x_{1}^{2} ) \right] \,,\nonumber \\
 p_{2}(x) & = & (1+\tau^2)^{-3}\left[ (1 + \tau^{6}) ( x_{1}^{6} + x_{2}^{6} +
  x_{3}^{6}) + 15\, \tau^{2} (x_{1}^{2} x_{2}^{4} + x_{2}^{2} x_{3}^{4} +
   x_{1}^{4} x_{3}^{2} ) \right.\nonumber\\
 & & \hspace{2em} + \left. 15\, \tau^{4} ( x_{1}^{4} x_{2}^{2} + x_{2}^{4}
x_{3}^{2} +
 + x_{1}^{2} x_{3}^{4} )\right] \,,\nonumber \\
 p_{3}(x) & = & x_{1}^{2} + x_{2}^{2} + x_{3}^{2} \,.
\end{eqnarray}

The essential elements of the associated $\widehat P$-matrix turn out to
be the following:

\begin{eqnarray}
 \widehat{P}_{11}(p)  &=&  192\, p_{1} p_{2} p_{3} + 336\, p_{1} p_{3}^{4}/5
+ 16\, p_{2}^{3}/9 - 1136\, p_{2}^{2} p_{3}^{3}/15 -
15968\, p_{2} p_{3}^{6}/75 \nonumber \\
  & &\hspace{2em}  + 81412\, p_{3}^{9}/1125 \,,\nonumber \\
\widehat{P}_{12}(p)  &=&  336\, p_{1} p_{3}^{2}/5 + 184\,p_{2}^{2} p_{3}/3 -
404\,p_{2} p_{3}^{4}/3 + 2656\,p_{3}^{7}/75 \,,\nonumber \\
 \widehat{P}_{22}(p) & =&  18 p_{1} - 12 p_{2} p_{3}^{2} + 174\,p_{3}^{5}/25
\,,
\end{eqnarray}
and, after the following MIBT:

\begin{eqnarray}
p'_{1} &=& (50625 p_1 - 123750 p_2 p_3^2 + 39285 p_3^5)/2 \,,\nonumber \\
p'_{2} &=& 225 p_2 - 93 p_3^3 \,,
\end{eqnarray}
one obtains, for $p'_3=1$, the matrix $R$ of class III.3($m=1$) reported in I:

\begin{eqnarray}
R'_{11}(\tilde p') &=& 1152 - 12 {p'}_1 - 168 {p'}_2 - 4 {p'}_1 {p'}_2 +
	   44 {p'}_2^2 + {p'}_2^3 \,,\nonumber \\
R'_{12}(\tilde p') &=& -6 {p'}_1 + 60 {p'}_2 + 5 {p'}_2^2  \,,\nonumber \\
R'_{22}(\tilde p') &=& 96 + {p'}_1 + 14 {p'}_2 \,,
\end{eqnarray}

\subsection{Type A$_4$}
The group acts on the plane $y_1+y_2+y_3+y_4+y_5=0$ of $\real^5$ by
permutations.  Therefore the group and its invariants can be obtained from the
reduction of the group S$_5$, acting on $y=(y_1, \dots, y_5)$ by permutations
of the coordinates.

A MIB for S$_5$ is yielded by $\{f_{5,1}(y), \dots , f_{5,5}(y)\}$, where
$f_{5,k}$ is defined in \eref{f}.  The reduction of the linear group $S_5$ can
be obtained by means of an orthogonal basis transformation in $\real^{5}$,
induced by the matrix

\begin{equation}
A \,=\, \left (  \begin{array}{ccccc}
  \frac{1}{\sqrt{2}} &      -\frac{1}{\sqrt{2}} &    0     &   0  & 0 \\[1em]
\frac{1}{\sqrt{6}} & \frac{1}{\sqrt{6}} & -\frac{2}{\sqrt{6}}&  0 & 0 \\ [1em]
\frac{1}{2 \sqrt{3}} &  \frac{1}{2 \sqrt{3}}& \frac{1}{2 \sqrt{3}}
					 &- \frac{3}{2 \sqrt{3}}& 0  \\ [1em]
\frac{1}{2 \sqrt{5}} & \frac{1}{2 \sqrt{5}} & \frac{1}{2 \sqrt{5}}
		     & \frac{1}{2 \sqrt{5}}& - \frac{4}{2 \sqrt{5}}  \\ [1em]
\frac{1}{\sqrt{5}}   & \frac{1}{\sqrt{5}} & \frac{1}{\sqrt{5}}
			   & \frac{1}{\sqrt{5}} & \frac{1}{\sqrt{5}}
\end{array} \right ) \,.
\end{equation}
The elements of the group of type A$_4$ are obtained as the principal minors
built with the first 4 rows and columns of the matrices $AgA^{-1},\ g\in S_5$
and, after setting $x=(x_1,\dots, x_4)$, a MIB is yielded by $\{p_1(x),
\dots p_4(x)\}$, where $p_a(x)$ is defined in \eref{pa}; explicitly:

\begin{eqnarray}
 p_1(x)  &=&  1800^{-1}\left(750\sqrt{6}x_1^4x_2 + 500\sqrt{6}x_1^2x_2^3 -
   250\sqrt{6}\,x_2^5 + 750\sqrt{3}\,x_1^4x_3 + 1500\sqrt{3}\,x_1^2x_2^2 x_3
	\right.\nonumber\\
& & \hspace{2em} + 750\sqrt{3}\,x_2^4x_3 + 750\sqrt{6}\,x_1^2x_2x_3^2
     - 250\sqrt{6}\,x_2^3x_3^2 + 250\sqrt{3}\,x_1^2x_3^3
     + 250\sqrt{3} \, x_2^2x_3^3  \nonumber\\
& & \hspace{2em} - 500\sqrt{3}\,x_3^5 + 450\sqrt{5}\,x_1^4x_4 +
900\sqrt{5}\,x_1^2x_2^2x_4
    + 450\sqrt{5}\,x_2^4x_4 + 900\sqrt{10}\,x_1^2x_2x_3x_4  \nonumber\\
& & \hspace{2em}  - 300\sqrt{10}\,x_2^3x_3x_4 + 450\sqrt{5}\,x_1^2x_3^2x_4
    + 450\sqrt{5}\,x_2^2x_3^2x_4 + 525\sqrt{5}\,x_3^4x_4 \nonumber\\
& & \hspace{2em} + 450\sqrt{6}\,x_1^2x_2x_4^2 - 150\sqrt{6}\,x_2^3x_4^2
   + 450\sqrt{3}\,x_1^2x_3x_4^2 + 450\sqrt{3}\,x_2^2x_3x_4^2\nonumber\\
& & \hspace{2em} - 300\sqrt{3}\,x_3^3x_4^2 + 90\sqrt{5}\,x_1^2x_4^3 +
90\sqrt{5}\,x_2^2x_4^3
    \left. 90\sqrt{5}\,x_3^2x_4^3 - 459\sqrt{5}\,x_4^5\right) \,,\nonumber\\
 p_2(x) & =&  60^{-1}\left(30\,x_1^4 + 60\,x_1^2x_2^2 + 30\,x_2^4 +
      60\sqrt{2}\,x_1^2x_2x_3 - 20\sqrt{2}\,x_2^3x_3 + 30\,x_1^2x_3^2
      + 30\,x_2^2x_3^2 \right. \nonumber\\
& & \hspace{2em} + 35\,x_3^4+ 12\sqrt{30}\,x_1^2x_2x_4 - 4\sqrt{30}\,x_2^3x_4
      + 121\sqrt{5}\,x_1^2x_3 x_4 + 121\sqrt{5}\,x_2^2x_3x_4 \nonumber\\
& & \hspace{2em} - 81\sqrt{5}\,x_3^3x_4 + \left.+ 18\,x_1^2x_4^2 +
18\,x_2^2x_4^2
     + 18\,x_3^2x_4^2 + 39\,x_4^4 \right)\,, \nonumber\\
 p_3(x) & = & 30^{-1}\left(15\sqrt{6}\,x_1^2x_2 - 5\sqrt{6}\,x_2^3 +
    15\sqrt{3}\,x_1^2x_3 + 15\sqrt{3}\,x_2^2x_3 - 10\sqrt{3}\,x_3^3 +
    9\sqrt{5}\,x_1^2 x_4      \right.\nonumber\\
& & \hspace{2em} + \left. 9\sqrt{5}\,x_2^2x_4 + 9\sqrt{5}\,x_3^2x_4 -
   9\sqrt{5}\,x_4^3\right)\,, \nonumber\\
 p_4(x) & = & x_1^2 + x_2^2 + x_3^2 + x_4^2\,.
\end{eqnarray}

The essential elements of the associated $\widehat{P}(p)$-matrix turn out
to be the following:

\begin{eqnarray}
\widehat{P}_{11}(p) &=& 5\left(12\, p_{2}^{2} + 128\, p_{1} p_{3} +
20\, p_{2} p_{4}^{2}  - 5 p_{4}^{4}\right)/48 \,,\nonumber \\
\widehat{P}_{12}(p) &=& \left(46\, p_{2} p_{3} + 84\, p_{1} p_{4} -
35\, p_{3} p_{4}^{2}\right)/6 \,,\nonumber \\
\widehat{P}_{13}(p) &=& \left( 40\, p_{3}^{2} + 66\, p_{2} p_{4}
- 15\, p_{4}^{3}\right)/8 \,,\nonumber \\
\widehat{P}_{22}(p) &=& 2\,\left(16\, p_{3}^{2} + 90\, p_{2} p_{4} -
15\, p_{4}^{3}\right)/15 \,,\nonumber \\
\widehat{P}_{23}(p) &=& 12\left(5\, p_{1} - p_{3} p_{4}\right)/5
\,,\nonumber \\
\widehat{P}_{33}(p) &=& 9\left(5\, p_{2} - p_{4}^{2}\right)/5 \,,
\end{eqnarray}
and, after the following MIBT:

\begin{eqnarray}
p'_{1} &=& 30\,\sqrt{15}\,\left(4\, p_1 - 3\,p_3 p_4\right)\,,\nonumber \\
p'_{2} &=& 3\,\left(20 p_2 - 7 p_4^2\right)  \,,\nonumber \\
p'_{3} &=& 6\,\sqrt{15}\,  p_3\, ,
\end{eqnarray}
one obtains, for $p'_3=1$, the matrix $R$ of class E$_{1}(s = 1)$, reported in
II:

\begin{eqnarray}
R'_{11}(\tilde p') &=& -4 {p'}_1 {p'}_3+3{p'}_2^2-27 {p'}_2+18
	(2 {p'}_3^2+9)\,,\nonumber \\
R'_{12}(\tilde p') &=& -6{p'}_1+{p'}_3 (5 {p'}_2+54)\,,\nonumber \\
R'_{13}(\tilde p') &=& 2 (3 {p'}_2+{p'}_3^2)\,,\nonumber \\
R'_{22}(\tilde p') &=& 3 {p'}_2+2 (4 {p'}_3^2+27)\,,\nonumber \\
R'_{23}(\tilde p') &=& {p'}_1+12 {p'}_3\,,\nonumber \\
R'_{33}(\tilde p') &=& {p'}_2+9\,.
\end{eqnarray}

\subsection{Type B$_4$}
The group acts on $x=(x_1, x_2, x_3, x_4)$ by permutations and sign changes
of the coordinates. A MIB can be chosen as follows:

\begin{equation}
\{p_a(x)  \,=\,  \summ 14i x_{i}^{10-2a}\}_{1\le a\le 4}\,.
\end{equation}
In this basis the essential elements of the $\widehat{P}(p)$ matrix turn
out to be the following:

\begin{eqnarray}
 \widehat{P}_{11}(p)  &=&  4\left (28 p_{1} p_{2} + 14 p_{2}
	p_{3}^{2} + 42 p_{1} p_{3} p_{4} -   21 p_{3}^{3} p_{4} -
     28 p_{2} p_{3} p_{4}^{2} + 14 p_{1} p_{4}^{3}
    + 7 p_{3}^{2} p_{4}^{3} - 14 p_{2} p_{4}^{4} \right.\nonumber \\
& & \hspace{2em} + \left. 7 p_{3} p_{4}^{5} - p_{4}^{7}\right)/3 \,,\nonumber
\\
 \widehat{P}_{12}(p) & =&  16 p_{2}^{2} + 36 p_{1} p_{3} - 6 p_{3}^{3} +
     36 p_{1} p_{4}^{2} - 18 p_{3}^{2} p_{4}^{2} - 32 p_{2} p_{4}^{3} +
     18 p_{3} p_{4}^{4} - 2 p_{4}^{6} \,,\nonumber \\
 \widehat{P}_{13}(p)  &=&   4\left(20 p_{2} p_{3} + 30 p_{1} p_{4}
     - 15 p_{3}^{2} p_{4} -  20 p_{2} p_{4}^{2} +  10 p_{3} p_{4}^{3} -
      p_{4}^{5}\right)/3 \,,\nonumber \\
 \widehat{P}_{22}(p)  &=&  3\left(20 p_{2} p_{3} + 30 p_{1} p_{4} -
       15 p_{3}^{2} p_{4} -   20 p_{2} p_{4}^{2} + 10 p_{3} p_{4}^{3} -
	 p_{4}^{5}\right)/2 \,,\nonumber \\
 \widehat{P}_{23}(p)  &=&  24 p_{1} \,,\nonumber \\
 \widehat{P}_{33}(p)  &=&  16 p_{2} \,,
\end{eqnarray}
and after the following MIBT:

\begin{eqnarray}
p'_{1} &=& 110592\, p_{1} - 55296\, p_{3}^{2} - 138240\, p_{2} p_{4}
	       + 98496\, p_{3} p_{4}^{2} - 15066\, p_{4}^{4}  \,,\nonumber \\
p'_{2} &=& 2304\, p_{2} - 2592\, p_{3} p_{4} + 612\, p_{4}^{3} \,,\nonumber\\
p'_{3} &=& 48\, p_{3} - 21\, p_{4}^{2} \,,
\end{eqnarray}
one obtains, for $p'_4=1$, the matrix $R$ of class E3$(s=1)$, reported in II:

\begin{eqnarray}
R'_{11}(\tilde p') &=& 2\, {p'}_1 ({p'}_2-54)+15\, {p'}_2^2+216\, {p'}_2
{p'}_3+ 324(4\,{p'}_3^2+18\, {p'}_3+81)\,,\nonumber\\
R'_{12}(\tilde p') &=& 6\, {p'}_1 {p'}_3-{p'}_2^2+18\, {p'}_2 ({p'}_3+12)+
1620\, {p'}_3\,,\nonumber\\
R'_{13}(\tilde p') &=& 6\, {p'}_1-{p'}_2 ({p'}_3-27)+54\, {p'}_3\,,\nonumber\\
R'_{22}(\tilde p') &=& 4 \left[3\, {p'}_1-{p'}_2 ({p'}_3+9)+27({p'}_3^2-
2\,{p'}_3+27)\right]\,,\nonumber\\
R'_{23}(\tilde p') &=& {p'}_1-3\, {p'}_2-6\, {p'}_3^2+108\,
{p'}_3\,,\nonumber\\
R'_{33}(\tilde p') &=& {p'}_2-12\, {p'}_3+81\,.
\end{eqnarray}

\subsection{Type D$_4$}
The group acts on $x=(x_1, x_2, x_3, x_4)$ by permutations and by changes of
an even number of signs of the coordinates. A MIB can be chosen as follows:

\begin{equation}
 p_{1}(x)  \,=\,  \summ 14i x_{i}^{6} \,,\qquad
    p_{2}(x)  \,=\,  \summ 14i x_{i}^{4} \,,\qquad
    p_{3}(x)  \,=\,  \prod_ {i = 1}^{4} x_{i} \,,\qquad
    p_{4}(x)  \,=\,  \summ 14i x_{i}^{2}\,.
\end{equation}

The essential elements of the associated $\widehat P$-matrix turn out to
be the following:

\begin{eqnarray}
\widehat{P}_{11}(p) &=& 6(5 p_{1} p_{2} - 30 p_{3}^{2} p_{4} +
5 p_{1} p_{4}^{2} - 5 p_{2} p_{4}^{3} +  p_{4}^{5}) \,,\nonumber \\
\widehat{P}_{12}(p) &=&   4 (3 p_{2}^{2} - 24 p_{3}^{2} + 8 p_{1} p_{4}
- 6 p_{2} p_{4}^{2} +  p_{4}^{4}) \,,\nonumber \\
\widehat{P}_{13}(p) &=&  6 p_{2} p_{3} \,,\nonumber \\
\widehat{P}_{22}(p) &=& 16 p_{1} \,,\nonumber \\
\widehat{P}_{23}(p) &=& 4 p_{3} p_{4}\,,\nonumber \\
\widehat{P}_{33}(p) &=& \left(2\, p_{1} - 3\, p_{2} p_{4} +
p_{4}^{3}\right)/6 \,,
\end{eqnarray}
and, after the following MIBT:

\begin{eqnarray}
p'_{1} &=& 12\left(12\, p_{1} - 15\, p_{2} p_{4} + 4\, p_{4}^{3}\right)\,,
	   \nonumber\\
p'_{2} &=& 48 \, \sqrt{3}\, p_{3}\,, \nonumber\\
p'_{3} &=& 6\,\left(- 2 p_{2} +  p_{4}^{2}\right)\,,
\end{eqnarray}
one obtains, for $p'_4=1$, the matrix $R$ of class E2$(s=1)$, reported in II:

\begin{eqnarray}
R'_{11}(\tilde p') &=& -4{p'}_1+5 {p'}_2^2+5 {p'}_3^2+12\,,\nonumber\\
R'_{12}(\tilde p') &=& 2 {p'}_2 ({p'}_3+3)\,,\nonumber\\
R'_{13}(\tilde p') &=& {p'}_2^2-{p'}_3 ({p'}_3-6)\,,\nonumber\\
R'_{22}(\tilde p') &=& {p'}_1+3 {p'}_3+6\,,\nonumber\\
R'_{23}(\tilde p') &=& 3 {p'}_2\,,\nonumber\\
R'_{33}(\tilde p') &=& {p'}_1-3 {p'}_3+6\,.
\end{eqnarray}

\subsection{Type F$_4$}

The group is the group generated by all the reflections in $\real^4$ which
leave invariant the hyperplanes $x_2-x_3=0$, $x_3-x_4=0$, $x_4=0$ and
$x_1-x_2-x_3-x_4=0$.

Let us define, for $ r= 2, 6, 8, 12 $:

\begin{eqnarray}
S_{k}(x) &=& \summ 1ki x_{i}^{k}  \,,\nonumber \\
I_{r}(x) &=& (8 - 2^{r - 1}) S_{r} + \summ 1{r/2 - 1}j \left(
  \begin{array}{c}
  r \\
  2 j
  \end{array}
  \right) S_{2 j} S_{r - 2 j} \nonumber\\
  &=& \sum_{1 \leq i < j \leq 4} [ (x_{j} + x_{i})^{r} + (x_{i} -
x_{j})^{r}]\,. \end{eqnarray}

Then, according to Metha, a MIB for the group can be defined to be the
following:

\begin{equation}
 p_{1}(x) \,=\,  I_{12}(x)\,, \qquad p_{2}(x)  \,=\,  I_{8}(x)\,,  \qquad
      p_{3}(x) \,=\,  I_{6}(x)\,,\qquad p_{4}(x)  \,=\,  I_{2}(x)/6\,,
\end{equation}
or, explicitely:

\begin{eqnarray}
 p_{1}(x)  &=&  924\, (x_{1}^{6} + x_{2}^{6} + x_{3}^{6} + x_{4}^{6})^{2} +
    990 (x_{1}^{4} + x_{2}^{4} + x_{3}^{4} + x_{4}^{4})
    (x_{1}^{8} + x_{2}^{8} + x_{3}^{8} + x_{4}^{8}) + 132 (x_{1}^{2}
       \nonumber \\
& & \hspace{2em}+ x_{2}^{2} + x_{3}^{2} + x_{4}^{2}) (x_{1}^{10} + x_{2}^{10} +
x_{3}^{10}
   + x_{4}^{10}) - 2040\, (x_{1}^{12} + x_{2}^{12}  + x_{3}^{12} + x_{4}^{12})
      \,,\nonumber \\
 p_{2}(x)  &=&  70\, (x_{1}^{4} + x_{2}^{4} + x_{3}^{4} + x_{4}^{4})^{2} +
    56 (x_{1}^{2} + x_{2}^{2} + x_{3}^{2} + x_{4}^{2})
    (x_{1}^{6} + x_{2}^{6} + x_{3}^{6} + x_{4}^{6}) -120 (x_{1}^{8}
       \nonumber \\
& & \hspace{2em}+ x_{2}^{8} + x_{3}^{8} + x_{4}^{8}) \,,\nonumber \\
 p_{3}(x) & =&  30\, (x_{1}^{2} + x_{2}^{2} + x_{3}^{2} + x_{4}^{2})
    (x_{1}^{4} + x_{2}^{4} + x_{3}^{4} + x_{4}^{4}) -
   24 (x_{1}^{6} + x_{2}^{6} + x_{3}^{6} + x_{4}^{6})  \,,\nonumber \\
 p_{4}(x) & =&  x_{1}^{2} + x_{2}^{2} + x_{3}^{2} + x_{4}^{2}\,.
\end{eqnarray}
The essential elements of the associated $\widehat{P}(p)$ matrix turn out to
be the following:

\begin{eqnarray}
 \widehat P_{11}(p)  &=&  \left(18711 p_2^2 p_3 + 1015740 p_1 p_2 p_4 -
     10178010 p_2 p_3^2 p_4
     - 625680 p_1 p_3 p_4^2 + 12496 p_3^3 p_4^2 \right. \nonumber\\
& & \hspace{2em} - 1675080 p_2^2 p_4^3 - 5264820 p_2 p_3 p_4^4 +
     24793560 p_1 p_4^5 + 14410440 p_3^2 p_4^5 \nonumber\\
& & \hspace{2em} -\left. 254481480 p_2 p_4^7 + 470719260 p_3 p_4^8 - 1709728560
     p_4^{11}\right)/810\,, \nonumber\\
 \widehat P_{12}(p)  &=&  4 \left(11970 p_1 p_3 - 610 p_3^3 + 49977
    p_2^2 p_4 - 175059 p_2 p_3 p_4^2 + 428220 p_1 p_4^3 + 110052 p_3^2
	p_4^3\right.       \nonumber\\
& & \hspace{2em}  - \left. 4455270 p_2 p_4^5
      + 9531630 p_3 p_4^6 - 32435640 p_4^9\right)/405\,, \nonumber\\
 \widehat P_{13}(p)  &=&  4 \left(243 p_2^2 + 2259 p_2 p_3 p_4
       + 15480 p_1 p_4^2 - 5572 p_3^2 p_4^2 - 129240 p_2 p_4^4
       + 278250 p_3 p_4^5 \right.\nonumber\\
& & \hspace{2em}   - \left. 847260 p_4^8\right)/45\,, \nonumber\\
 \widehat P_{22}(p)  &=&  16 \left(21 p_2 p_3 + 84 p_1 p_4 - 28 p_3^2
   p_4 - 840 p_2 p_4^3 + 1890 p_3 p_4^4 - 6120 p_4^7\right) / 3\,,
       \nonumber\\
 \widehat P_{23}(p)  &=&  32 \left(18 p_1 + 7 p_3^2 - 63 p_2 p_4^2
       + 273 p_3 p_4^3 - 1134 p_4^6\right) / 9\,, \nonumber\\
 \widehat P_{33}(p)  &=&  72 \left(-12 + p_2 p_4 + 4 p_3 p_4^2\right)
\end{eqnarray}
and, after the following MIBT:

\begin{eqnarray}
 p'_1   &=&  96\, \sqrt {6}\, \left(288\, p_1 - 77\, p_3^2 -
       3762\, p_2 p_4^2 + 8832\, p_3 p_4^3 - 29511\, p_4^6\right)/5\,,
\nonumber\\
 p'_2  & = &  108\, \left(16\, p_2 - 56\, p_3 p_4 + 255\, p_4^4\right)/5\,,
\nonumber\\
 p'_3   &=&  6\, \sqrt {6}\, \left(2\, p_3 - 15\, p_4^3\right)\,,
\end{eqnarray}
one obtains, for $p'_4=1$, the matrix $R$  of class E4$(s=1)$, reported in II:

\begin{eqnarray}
R'_{11}(\tilde p') &=& -6 \left[12 {p'}_1 {p'}_3-21 {p'}_2^2-{p'}_2
(11 {p'}_3^2-864)-36 (17 {p'}_3^2+97 2)\right]\,,\nonumber\\
R'_{12}(\tilde p') &=& -6 \left[12 {p'}_1-{p'}_3 (21
{p'}_2+{p'}_3^2+756)\right]\,,\nonumber\\
R'_{13}(\tilde p') &=& {p'}_2^2+180 {p'}_2+60 {p'}_3^2\,,\nonumber\\
R'_{22}(\tilde p') &=& 6(9 {p'}_2+7 {p'}_3^2+972)\,,\nonumber\\
R'_{23}(\tilde p') &=& {p'}_1+180 {p'}_3\,,\nonumber\\
R'_{33}(\tilde p') &=& 5 {p'}_2+324\,.
\end{eqnarray}

\subsection{Type H$_4$}
The group is defined as the group generated by all the reflections in
$\real^4$ which leave invariant the hyperplanes $x_3=0$,  $x_4=0$,
$\tau^{-1}x_2-\tau x_3-x_4=0$ and  $\tau^{-1}x_1-\tau x_2-x_4=0$, where
$\tau$ is defined in \eref{tau}.

For $j, k, l, m$ in the set $\{1, 2, 3, 4\}$, let us define the following
symbols:

\begin{equation}
\eta_{jk} \,=\, \cases{-1, & for $j=k$ ;\cr \ 1, & otherwise; \cr}
\end{equation}
and the following expressions:
\begin{equation}
\xi(m, j, k, l;x) \,=\, \tau x_j \eta_{mj} + \tau^{-1} x_k \eta_{mk} +
	 x_l\eta_{ml}\,;
\end{equation}

\begin{equation}
 \chi(j, k, l, n;x) \,=\, \xi(0, j, k, l;x)^{2n} + \xi(j, j, k, l;x)^{2n} +
		  \xi(k, j, k, l;x)^{2n} + \xi(l, j, k, l;x)^{2n}\,;
\end{equation}
\begin{equation}
\psi(j, n;x) \,=\, \left(\summ 14k \eta_{jk} x_k\right)^{2n}.
\end{equation}

\begin{eqnarray}
 I_n(x) &= & \summ 14k (2x_k)^{2n} +  \summ 04k \psi(k;x) + \chi(1,2,3,n;x)
+         \chi(1,3,4,n;x)\nonumber\\
& & \hspace{2em} + \chi(1,4,2,n;x)+ \chi(2,4,3,n;x).
\end{eqnarray}

Then, according to Metha, a MIB for $H_4$ can be chosen in the following
way:
\eq{ p_1(x)\, =\, I_{30}(x),\quad p_2(x) \,=\, I_{20}(x),\quad p_3(x) \,=\,
I_{12}(x),\quad p_4(x) \,=\, \summ 14i x_i^2\,.}
The essential elements of the associated $\widehat P$-matrix turn out to be
the following:

\begin{eqnarray}
 \widehat P_{11}(p)  & = &
      53911 \,p_2 p_3^3 p_4/7616 + 7051785\, p_1 p_3^2 p_4^2/56
      + 4821334245\, p_2^2 p_3 p_4^3/25432 \nonumber \\
& & \hspace{2em}  + 2186873325\, p_1 p_2 p_4^4/34 -  645826368707\, p_3^4
p_4^5/150528
       \nonumber \\
& & \hspace{2em}  - 116309076672555 \,p_2 p_3^2 p_4^7/23936 -  211651127025
\,p_1 p_3
	p_4^8/2 \nonumber \\
& & \hspace{2em} - 21238646708813625 \,p_2^2 p_4^9/18496 - 283066493617380915
\,p_3^3
    p_4^{11}/9856 \nonumber \\
& & \hspace{2em}  - 2811241304150172075 \,p_2 p_3 p_4^{13}/544 +
2218140033302250 \,p_1
      p_4^{14} \nonumber \\
& & \hspace{2em}  + 7676790020731375739325 p_3^2 p_4^{17}/224
   -  15228773425368084479625 p_2 p_4^{19}/136 \nonumber \\
& & \hspace{2em}    + 253639342346876415408375 \,p_3 p_4^{23}/8 \nonumber \\
& & \hspace{2em} - 8927280781972196041680013125 \,p_4^{29}/4\,, \nonumber \\
 \widehat P_{12}(p) & = &
   26741 \,p_3^4/61740 + 2357459 \,p_2 p_3^2 p_4^2/56 + 1473900 \,p_1 p_3
      p_4^3+ 482714505 \,p_2^2 p_4^4/22  \nonumber \\
& & \hspace{2em} - 1869793842241 \,p_3^3 p_4^6/7056 - 7474081897815 \,p_2 p_3
p_4^8/44
	  \nonumber \\
& & \hspace{2em} + 327790212400 \,p_1 p_4^9 + 66300758108151125 \,p_3^2
p_4^{12}/308
	\nonumber \\
& & \hspace{2em} - 34964650402339275 \,p_2 p_4^{14} + 128303960304363056775
\,p_3 p_4^{18}
	\nonumber \\
& & \hspace{2em} - 993168612785995074523500 \,p_4^{24}\,, \nonumber \\
 \widehat P_{13}(p)  &= &
   429975 \,p_2^2/3179 + 4236791 \,p_3^3 p_4^2/168 + 2490103980 \,p_2 p_3
       p_4^4/187 \nonumber \\
& & \hspace{2em}     + 92563800 p_1 p_4^5 - 7157961621135 p_3^2 p_4^8/88 -
330008206170825
	p_2 p_4^{10}/34 \nonumber \\
& & \hspace{2em}   + 36082213228297575 \,p_3 p_4^{14} - 271058613666147798750
\,p_4^{20}\,,
	\nonumber \\
 \widehat P_{22}(p) & =&
   2657644 \,p_3^3 p_4/138915 + 24894256 \,p_2 p_3 p_4^3/21 + 5074562560 \,p_1
       p_4^4/819 \nonumber \\
& & \hspace{2em}   - 1343036354024 \,p_3^2 p_4^7/441 - 644361278880 \,p_2 p_4^9
\nonumber \\
& & \hspace{2em}   + 49357102179756800 \,p_3 p_4^{13}/21 -
4855838749013355171200 p_4^{19}
       /273\,, \nonumber \\
 \widehat P_{23}(p) & = &
   10880 \,p_1/13 + 1322362 \,p_3^2 p_4^3/21 - 80135160 \,p_2 p_4^5
 + 296137692080 \,p_3 p_4^9 \nonumber\\
& & \hspace{2em} - 29473265236173600 \,p_4^{15}/13\,, \nonumber \\
 \widehat P_{33}(p) & = &
       5040 p_4 (7 \,p_2 - 18224 \,p_3 p_4^4 + 167747160 \,p_4^{10})/17
\end{eqnarray}
and, after the following MIBT:

\begin{eqnarray}
 p'_1  &= & 98415000\,\left( 576\, p_1/1001 - 108555\, p_3^2 p_4^3/539 -
   11535372\, p_2 p_4^5/187 \right.\nonumber\\
  & & \hspace{2em}  \left. + 17469633928\, p_3 p_4^9/77 - 1724135397013808\,
p_4^{15}/1001
      \right)\,, \nonumber\\
 p'_2  &=&  2187000\left( -6\, p_2/187 + 1307\, p_3 p_4^4/11 -
       908706\, p_4^{10} \right)\,,\nonumber\\
 p'_3  &=&  1620\left( -p_3/7 + 1130\, p_4^6\right) \,.
\end{eqnarray}
one obtains, for $p'_4=1$, the matrix $R$ of class E5$(s=1)$ reported in
II\footnote{In II, the sign of the monomial $p_1p_3$ in the expression
of $R'_{11}$ is wrong and should be changed; accordingly should be
changed the expression of $A(p)$.}:

\begin{eqnarray}
 \widehat R'_{11}(p') & = & -36\,p_2^2(-25380 + 19\,p_3) +
      90\,p_1(-1166400 + 12\,p_2 + 1080\,p_3 + p_3^2) \nonumber \\
& & \hspace{2em}    - p_2(-75582720000 - 342921600\,p_3 - 6480\,p_3^2 +
29\,p_3^3) \nonumber\\
& & \hspace{2em}   + 45\,(918330048000000 + 906992640000\,p_3 +
1102248000\,p_3^2 \nonumber \\
& & \hspace{2em}     - 339120\,p_3^3 + 209\,p_3^4)\,, \nonumber \\
 \widehat R'_{12}(p')  &=&  -486\,p_2^2 + 360\,p_1(540 + p_3) -
     9\,p_2(-34992000 - 3240\,p_3 + 19\,p_3^2) \nonumber \\
& & \hspace{2em}   - p_3\,(-89754480000 + 28431000\,p_3 - 45360\,p_3^2 +
p_3^3)\,,\nonumber \\
 \widehat R'_{13}(p') & =&  2160\,p_1 + p_2^2 - 1980\,p_2(-1080 + p_3)
	- 55\,(-4860 + p_3)p_3^2\,, \nonumber \\
 \widehat R'_{22}(p') & =&  212576400000 + 540\,p_1 - 45927000\,p_3 +
       218700\,p_3^2 - 19\,p_3^3 - 324\,p_2(2025 + p_3)\,, \nonumber \\
 \widehat R'_{23}(p')  &=&  p_1 - 810\,p_2 - 495\,(-2700 + p_3)p_3\,,
	\nonumber \\
 \widehat R'_{33}(p')  &=&  11\,p_2 - 6750\,(-1296 + p_3)\,.
\end{eqnarray}

\section{Reducible reflection groups with 2, 3 and 4 basic invariants}

In this section we shall state the rules for building the $\widehat
P$-matrices of a reducible coregular linear group $G$, in a standard A-basis,
starting from the $\widehat P$-matrices associated to its irreducible
components. This will allow us, in particular, to
derive the explicit form of the $\widehat P$-matrices
of all the reducible reflection groups whose orbit spaces have dimensions $\le
4$ and state which of the allowable $\widehat P$-matrices determined in I and
II are related to these groups.

Let $G^{(1)}$ and $G^{(2)}$ be irreducible CLG's, acting, respectively, in
$\real^{n_1}$ and $\real^{n_2}$.  Then, the set of matrices

\eq{G\,=\,\{g^{(1)}\oplus g^{(2)}\}_{g^{(\alpha)}\in G^{(\alpha)}\atop
\alpha =1,2}}
forms a coregular linear group, acting on the vectors
$x=x^{(1)}\oplus x^{(2)},\ x\in\real^{n_1+n_2}=\real^{n_1}\oplus \real^{n_2}$.
If the groups $G^{(\alpha)},\ \alpha =1,2$ are generated by reflections,
$G$ is a reflection group isomorphic to $G_1\otimes G_2$.

Let us denote by $p_i^{(\alpha)}(x^{(\alpha)}),\ i=1,\dots ,q_\alpha,\
x^{(\alpha)}\in \real^{n_\alpha}$
the elements of a standard MIB relative to $G^{(\alpha)},\ \alpha=1,2$ and by
$\{d_i^{(\alpha)}\},\ \widehat P^{(\alpha)}(p^{(\alpha)})$ the associated set
of degrees and $\widehat P$-matrix.  We shall assume $d_1^{(1)}\ge d_1^{(2)}$.

A set of basic polynomial invariants of $G$ is yielded by

\eq{p^{(+)}(x)\,=\,\{p^{(1)}(x^{(1)}),p^{(2)}(x^{(2)})\}\,.}
and the associated $\widehat P$-matrix has the following form:

\eq{ \widehat P^{(+)}(p^{(+)})\,=\,\widehat P^{(1)}(p^{(1)})\oplus \widehat
P^{(2)}(p^{(2)})\,.}

If $\{p^{(\alpha)}\}$ is a standard $A$-basis relative to $G^{(\alpha)}$,
$A^{(\alpha)}(p^{(\alpha)})$ is the complete active factor of det$\widehat
P^{(\alpha)}(p^{(\alpha)})$ and $w^{(\alpha)}$ is its weight, then

\eq{A^{(+)}(p^{(+)})\,=\,A^{(1)}(p^{(1)})A^{(2)}(p^{(2)})}
is the complete active factor of det$\widehat P^{(+)}$; it satisfies the
following relations:

\eq{ \summ 1{q_1+q_2}b\widehat P^{(+)}_{ab}(p^{(+)})\partial_b
A^{(+)}(p^{(+)})\,=\,\lambda^{(+)}_a A^{(+)}(p^{(+)})\,,\qquad a=1,\dots
,q_1+q_2\,,}
where
\eq{{\lambda^{(+)}}\,=\,\lambda^{(1)}\oplus \lambda^{(2)}\,=\,(0,\dots , 0,
2w^{(1)}, 0, \dots , 0, 2w^{(2)})\,.}
Therefore, a standard $A$-basis $\{p\}$ of $G$ can be obtained from
$\{p^{(+)}\}$ by means of a basis transformation $p=f(p^{(+)})$, with
$f(p^{(+)})$ defined, for instance, by the following relations:

\eqll{
\begin{array}{rcl}
p_k &=& p_{i(k)}^{(+)},\qquad\qquad k=1,\dots ,q_1+q_2-2\,,\\
p_{q_1+q_2-1} &=& b_{q_1+q_2-1}\left(w^{(2)}p_{q_1}^{(1)} -
w^{(1)}p_{q_2}^{(2)}\right)\,, \\
p_{q_1+q_2} &=& p_{q_1}^{(1)}+p_{q_2}^{(2)}\,;
\end{array}}{103}
in \eref{103}, the set $\{i(1),\dots ,i(q_1+q_2-2)\}$ is a permutation of
the indices $\{1,\dots , q_1+q_2-2\}$,
such that the degrees of the invariants $p_k$ are non increasing functions of
$k$.  The parameters $b_k$ are arbitrary and will be chosen so that the
$\widehat P$-matrices can be easily compared with the results reported in I
and II.

The $\widehat P$-matrix associated to the MIB $\{p\}$ is determined by the
following relation (see \eref{equiv}):

\eq{\widehat P(p)\,=\,\left. J(p^{(+)})\, \widehat P^{(+)}(p^{(+)})\,
J^T(p^{(+)})\right|_{p^{(+)}\,=\,f^{-1}(p)},}
where $J(p^{(+)})$ denotes the Jacobian matrix

\eq{J_{ab}(p^{(+)})\,=\,\frac{\partial f_a(p^{(+)})}{\partial p_b^{(+)}},
\qquad a,b= 1,\dots ,q_1+q_2.}

In the aim of determining the $\widehat P$-matrices of all the reducible
reflection groups whose orbit spaces have
dimensions 3 and 4, let us now specialize the construction we have
described to the cases $q_1=2,\ q_2=1,2$ and to each of the different
cases one can get starting from $q_1=3,\ q_2=1$.
We shall denote by $w$ the weight of the complete active factor $A(p)$ of
det$\widehat P(p)$:

\eq{w\,=\,w^{(+)} \,=\,w^{(1)}+ w^{(2)}\,.}

Below, for each of the different reducible groups that can be obtained in
this way, we shall list the values of the degrees, the explicit form of the
matrix $ R^{(+)}$, the MIBT leading to a standard $A$-basis and the
transformed form of the matrix $R(p)$, evaluated at $p_q=1$, to be compared
with the results of I and II.

\vskip5truemm
\subsection{Case $q_1=2,\ q_2=1$}

\eq{
\begin{array}{lll}
\begin{array}{l}
d_1\,=\,d_1^{(1)}\,=\,m+1\,,\\
d_2\,=\,d_2^{(1)}\,=\,2\,,
\end{array}
&\qquad
\begin{array}{l}
d_3\,=\,d_1^{(2)}\,=\,2\,,\\
w  \,=\,2d_1^{(1)}+2\,=\,2m+4\,;
\end{array}
&\qquad
m\in\natural_*;
\end{array}
}
\vskip3truemm

\eq{
\begin{array}{ll}
\begin{array}{rcl}
R_{11}^{(+)}(p^{(+)}) &=& {p_{2}^{(1)}}^m\,,\\
R_{12}^{(+)}(p^{(+)}) &=& p_{1}^{(1)}\,,\\
R_{22}^{(+)}(p^{(+)}) &=& p_{2}^{(1)}\,,
\end{array}
&\qquad
\begin{array}{rcll}
R_{a3}^{(+)}(p^{(+)}) &=& 0\,,\quad a=1,2,\\
R_{33}^{(+)}(p^{(+)}) &=& p_1^{(2)}\,,\\
R_{44}^{(+)}(p^{(+)}) &=& p_2^{(2)}\,;
\end{array}
\end{array}
}
\vskip3truemm

\eq{
\begin{array}{lcl}
p_{1} &=& (2+m)^{m/2}p_{1}^{(1)}\,,\\
p_{2} &=& -p_{2}^{(1)}+(m+1)p_{1}^{(2)} \,, \\
p_{3} &=& p_{2}^{(1)}+p_{1}^{(2)}\,;
\end{array}
}
\vskip3truemm

\eq{
\begin{array}{ll}
\begin{array}{rcl}
R_{11}(p) &=& \left[(1 + m)p_3 - p_2\right]^m\,,\\
R_{12}(p) &=& -p_1\,;\\
\end{array}
& \qquad
\begin{array}{rcl}
R_{22}(p) &=& (1 + m)p_3 + mp_2\,,\\
R_{a3}(p) &=& p_a,\qquad a=1,2,3\,;
\end{array}
\end{array}
}
which, for $p_3=1$, is the class II $R$-matrix reported in II.

\vskip5truemm
\subsection{Case $q_1=2,\ q_2=2$}

\eq{
\begin{array}{lll}
\begin{array}{l}
d_1\,=\,d_1^{(1)}\,=\,j_1+1\,,    \\
d_2\,=\,d_1^{(2)}\,=\,j_2+1\,,
\end{array}
&
\begin{array}{l}
d_3\,=\,d_2^{(1)}\,=\,2\,,         \\
d_4\,=\,d_2^{(2)}\,=\,2\,,
\end{array}
&
\begin{array}{l}
j_1,j_2\,\in \,\natural_* \,;\\
\end{array}
\end{array}
}

\eq{w\,=\,w^{(1)}+w^{(2)}\,=\,2(j_1+j_2+2)\,; }
\vskip3truemm

\eq{
\begin{array}{l}
\begin{array}{lll}
\begin{array}{lll}
R_{11}^{(+)}(p^{(+)}) &=& {p_{2}^{(1)}}^{j_1}\,,\\
R_{12}^{(+)}(p^{(+)}) &=& p_{1}^{(1)}\,,\\
\end{array}
&\
\begin{array}{l}
R_{22}^{(+)}(p^{(+)}) \,=\, p_{2}^{(1)}\,,  \\
R_{33}^{(+)}(p^{(+)}) \,=\,  {p_{2}^{(2)}}^{j_2}\,,
\end{array}
&\
\begin{array}{l}
R_{34}^{(+)}(p^{(+)}) \,=\,  p_{1}^{(2)}\,,\\
R_{44}^{(+)}(p^{(+)}) \,=\, p_{2}^{(2)}\,;
\end{array}
\end{array}    \\
\begin{array}{l}
\ R_{ab}^{(+)}(p^{(+)}) \,=\, 0\,, \qquad a=1,2\,,\quad  b=3,4\,;
\end{array}
\end{array}
}
\vskip3truemm

\eq{
\begin{array}{ll}
\begin{array}{lll}
p_{1}&=&(j_1+j_2+2)^{\frac {j_1}2}p_{1}^{(1)}\,,\\
p_{2}&=&(j_1+j_2+2)^{\frac{j_1}2}p_{1}^{(2)}\,, \\
\end{array}
&\qquad
\begin{array}{lll}
p_{3} &=& (j_2+1)p_{2}-(j_1+1)p_{4}\,, \\
p_{4} &=& p_{2}^{(1)}+p_{2}^{(2)}\,;
\end{array}
\end{array}
}
\vskip3truemm

\eq{
\begin{array}{ll}
\begin{array}{l}
R_{11}(p) \,=\, (j_1+j_2+2)\left[- p_3 + (j_1+1) p_4\right]^{j_1}\,,\\
R_{22}(p) \,=\, (j_1+j_2+2)\left[p_3+(j_2+1)p_4\right]^{j_2}\,,\\
R_{33}(p) \,=\, (j_1+1)(j_2+1)+(j_1-j_2)p_3\,,
\end{array}
&\quad
\begin{array}{l}
R_{12}(p) \,=\, 0\,,          \\
R_{13}(p) \,=\, -(j_2+1)p_1\,,  \\
R_{23}(p) \,=\,   (j_1+1) p_2\,,
\end{array}
\end{array}
}
which, for $p_3=1$, is the class A8($j_1,j_2$) $R$-matrix reported in II.
For $j_i=1$ the group $G^{(i)}\simeq Z_2\otimes Z_2$ is reducible.
\vskip5truemm
\subsection{Cases $q_1=3,\ q_2=1$}

\eq{
\begin{array}{lll}
\begin{array}{rcl}
d_1&=&d_1^{(1)}\,,\\
d_2&=&d_2^{(1)}\,,
\end{array}
&\qquad
\begin{array}{rcl}
d_3&=&d_3^{(1)}\,,\\
d_4&=&d_1^{(2)}\,,
\end{array}
&\qquad
w\,=\,w^{(1)}+2\,;
\end{array}
}
\vskip3truemm

\eq{\begin{array}{rcll}
R^{(+)}(p^{(+)})_{ab} &=& R_{ab}^{(1)}(p^{(1)})\,,&a,b=1,2,3,\\
R^{(+)}(p^{(+)})_{a4} &=& 0\,,                    &a=1,2,3,             \\
R^{(+)}(p^{(+)})_{44} &=& R_{11}^{(2)}(p^{(2)})\,=\,p_1^{(2)}\,;&
\end{array}
}
\vskip3truemm

\eq{
\begin{array}{ll}
\begin{array}{rcl}
p_1 &=& b_1p_1^{(1)}\,,\\
p_2 &=& b_2p_2^{(1)}\,,\\
\end{array}
&\qquad
\begin{array}{rcl}
p_3 &=& b_3\left(p_3^{(1)}-w^{(1)}p_1^{(2)}/2\right)\,,  \\
p_4 &=& p_3^{(1)}+p_1^{(2)}\,.
\end{array}
\end{array}
}

The matrix $R^{(1)}$ can be chosen from 3 different classes, denoted as
III.1, III.2, III.3 in II and corresponding, respectively, to the groups
of type A$_3$ (or D$_3$), B$_3$ and H$_3$.

\vskip5truemm
Choosing as $G^{(1)}$ a \underline{type A$_3$} group one obtains:

\eq{
\begin{array}{ll}
\begin{array}{lll}
d_1 &=& 4\,,        \\
d_2 &=& 3\,,
\end{array}
&\qquad
\begin{array}{lll}
d_3     &=& d_4\,=\,2\,,\\
w^{(1)} &=& 12\,;
\end{array}
\end{array}
}
\vskip3truemm

\eq{
\begin{array}{ll}
\begin{array}{rcl}
p_1 &=& 49p_1^{(1)}\,,                         ,\\
p_2 &=& -7^{\frac 32}p_2^{(1)}\,;
\end{array}
&\qquad
\begin{array}{rcl}
p_3 &=& p_3^{(1)}-6p_1^{(2)}\,,\\
p_4 &=& p_3^{(1)}+p_1^{(2)}\,;
\end{array}
\end{array}
}
\vskip3truemm

\begin{eqnarray}
R_{11}(p) &=& 7 \left[p_2^2 - p_1(p_3 + 6p_4) + 2(p_3 + 6p_4)^{3}\right]\,,
	   \nonumber\\
R_{22}(p) &=& 7\left[p_1 + 2(p_3 + 6p_4)^{2}\right]\,,   \nonumber \\
R_{12}(p) &=& 14p_2(p_3 + 6p_4)\,,\nonumber\\
R_{13}(p) &=& p_1, \qquad R_{23}(p) \,=\, p_2\,, \qquad
			      R_{33}(p) \,=\, - 5p_3 + 6p_4\,,
\end{eqnarray}
which, for $p_4=1$, is the class B4($s=1$) $R$-matrix reported in II.

\vskip5truemm
Choosing as $G^{(1)}$ a \underline{type B$_3$} group one obtains:

\eq{
\begin{array}{ll}
\begin{array}{rcl}
d_1 &=& 6\,,      \\
d_2 &=& 4\,,
\end{array}
&\qquad
\begin{array}{rcl}
d_3     &=& d_4\,=\,2\,, \\
w^{(1)} &=& 18\,;
\end{array}
\end{array}
}
\vskip3truemm

\eq{
\begin{array}{lll}
\begin{array}{lll}
p_1 &=& 10^{3}p_1^{(1)}\,,\\
p_2 &=& -10^{2}p_2^{(1)}\,,
\end{array}
&\qquad
\begin{array}{rcl}
p_3 &=& p_3^{(1)}-9p_1^{(2)}\,,\\
p_4 &=& p_3^{(1)}+p_1^{(2)}\,;
\end{array}
\end{array}
}
\vskip3truemm

\begin{eqnarray}
 R_{11}(p)  &=&  10\left\{p_1\left[p_2 - 4(p_3 + 9p_4)^{2}\right] + 8(p_3 +
	      9p_4) \left[p_2^2 \right.\right. \nonumber\\
  & & \hspace{2em}  \left.\left.\qquad\quad + 2p_2 (p_3 + 9p_4)^2 + 8(p_3 +
9p_4)^4 \right]
			\right\}\,,\nonumber\\
 R_{12}(p)  &=&  10\left\{2p_1(p_3 + 9p_4) - p_2\left[p_2 - 12(p_3 +
	    9p_4)^{2} \right]\right\}\,, \nonumber\\
 R_{22}(p)  &=&  10 \left\{p_1 - 4(p_3 + 9p_4)\left[p_2 - 4(p_3 +
	 9p_4)^{2}\right] \right\}\,, \nonumber\\
 R_{13}(p)  &=&  p_1, \qquad R_{23}(p) \,=\, p_2\,, \qquad
			      R_{33}(p) \,=\, - 8p_3 + 9p_4\,,
\end{eqnarray}
which, for $p_4=1$, is the class C6($1$) $R$-matrix reported in II.

\vskip5truemm
Choosing as $G^{(1)}$ a \underline{type H$_3$} group one obtains:

\eq{
\begin{array}{ll}
\begin{array}{lll}
d_1 &=& 10\,,       \\
d_2 &=& 6\,,
\end{array}
&\qquad
\begin{array}{lll}
d_3     &=& d_4\,=\,2\,,   \\
w^{(1)} &=& 30\,,
\end{array}
\end{array}
}
\vskip3truemm

\eq{
\begin{array}{rcl}
\begin{array}{rcl}
p_1 &=& 16^{5}p_1^{(1)}\,,\\
p_2 &=& -16^{3}p_2^{(1)}\,,
\end{array}
&\qquad
\begin{array}{rcl}
p_3 &=& p_3^{(1)}-15p_1^{(2)}\,,\\
p_4 &=& p_3^{(1)}+p_1^{(2)}\,;
\end{array}
\end{array}
}
\vskip3truemm

\begin{eqnarray}
 R_{11}(p)  &= & 16 \left\{4p_1 (p_3 + 15p_4)\left[p_2 - 3(p_3 +
	15p_4)^{3}\right] - \left[p_2 - 48(p_3 + \right.\right.\\ \nonumber
  & & \hspace{2em} \left.\left. 15p_4)^{3}\right] \left[p_2^2 + 4p_2 (p_3 +
15p_4)^{3} +
	+ 24(p_3 + 15p_4)^{6}\right] \right\}\,, \nonumber\\
 R_{12}(p) & = & 16(p_3 + 15j_1p_4)\left[-5p_2^2 + 6p_1(p_3 + 15p_4)+
       60p_2(p_3 + 15p_4)^{3}\right]\,, \nonumber\\
 R_{22}(p)  &=&  16\left[p_1 - 14p_2(p_3 + 15p_4)^{2} + 96(p_3 +
       15p_4)^{5}\right] \,,\nonumber \\
 R_{13}(p) & =&  p_1,\qquad R_{23}(p) \,=\, p_2,\qquad
	R_{33}(p) \,=\, - 14p_3 + 15p_4\,,
\end{eqnarray}
which, for $p_4=1$, is the class D4($j_1$) $R$-matrix reported in II.

\section{Concluding remarks on the mathematical results}
Starting from explicit forms for the basic polynomial invariants of the finite
coregular groups, that can be found in the mathematical literature, we have
computed the associated $\widehat P(p)$ matrices.  The equalities and
inequalities, defining the orbit spaces of the groups as semi-algebraic
sub-varieties of $\real^q$, can be easily obtained as semi-positivity
conditions on these matrices.

The computation has been limited to the groups with less than 5 basic
polynomial invariants, since the main aim of our work was to test the
completeness of the allowable solutions of the canonical equation listed in I
and II and to find the corresponding generating groups.  The test has been
positive:  all the $\widehat P$-matrices generated by finite coregular groups
appear in the lists of allowable $\widehat P$-matrices reported in I and II.
In particular:

\begin{enumerate}
\item For $q=2$ all the allowable $\widehat P$-matrices are generated by
coregular {\em finite} groups; the case $q=2$ is exceptional, since the
canonical equation puts no restrictions on the allowable $\widehat
P$-matrices.  On the contrary, for $q=3,4$, only the fundamental elements
(scale parameter $s=1$) of some towers of allowable $\widehat P$-matrices are
generated by coregular {\em finite} groups; in this case, the existence of
towers of solutions of the canonical equation has probably no group
theoretical meaning, but is only an artifact related to an invariance of the
canonical equation under scaling of the degrees of the basic polynomial
invariants. By now, we cannot exclude however that the higher
elements in the towers are generated by non-finite groups.

\item For $q=2,3,4$ all the {\em fully active} allowable solutions of the
canonical equation are generated by {\em finite} coregular linear groups; the
{\em irreducible} ones are associated to {\em irreducible} linear groups.
\end{enumerate}

As for the fundamental allowable $\widehat P$-matrices for which we have not
found a finite coregular generating group, various, more or less obviuos,
interpretations are possible; they might be generated by

\begin{itemize}
\item[i)] non-finite compact coregular groups,
\item[ii)] non-coregular groups,
\item[iii)] non-minimal integrity bases of finite or non-finite compact
coregular groups,
\item[iv)] direct sums of non-fundamental allowable $\widehat P$-matrices.
\end{itemize}
This is probably an incomplete list.

We shall try to clarify this point in two forthcoming papers.
The correspondence between the classification of the allowable $\widehat
P$-matrices determined in I and II and the generating finite reflection groups
with less than 5 basic polynomial invariants is summarized in Tables 1-4,
where by a group of type I$_2(1)$ we mean the group
Z$_2\otimes$Z$_2$, (Z$_2=\{\pm 1\}$).  The case $q=1$ is trivial, as there is
only one allowable $\widehat P$-matrix generated by the group Z$_2$.  For
$q=2$, for each choice of the degree $d_1$ there is only one equivalence class
of allowable $\widehat P$-matrices, which are generated by at least one finite
coregular linear group.  For $q=3$, all the fundamental elements in each tower
of (classes of) allowable solutions are generated by at least one finite
coregular group, but for the $\widehat P$-matrices of class I.  For $q=4$, the
number of (classes of) reducible allowable solutions which are not generated
by at least one finite coregular group is much higher.

\section{Physical applications. An example}

The $\widehat P$ matrix approach to the study of orbit spaces has been, or can
be used, in various physical contexts, where the study of covariant functions
is important, as already stressed in the introduction.  Typical examples are
the determination of patterns of spontaneous symmetry and/or supersymmetry
breaking \cite{020,220} in gauge field theories of elementary particles, the
analysis of phase spaces and structural phase transitions in the framework of
Landau's theory \cite{470,471} and in cosmology (phase transitions in the hot
Universe \cite{Linde}).  Applications can be done in covariant bifurcation
theory \cite{231} or in crystal field theory and in most areas of solid state
theory where use is made of symmetry adapted functions.

Most of the groups dealt with in the preceding sections are crystallographic
groups, they are therefore symmetry groups of regular polyhedra in 2, 3 or 4
dimensions and of root diagrams of simple Lie algebras.  In particular,
I$_2(m)$ denotes, in Coxeter's notation, the dihedral group denoted by
C$_{n{\rm v}}$ in the standard physical notation \cite{470},
I$_2(m)\otimes\integer_2$ denotes $D_{n{\rm h}}$, while D$_3$,\ B$_3$ and H$_3$
correspond, respectively, to the groups T$_{\rm d}$, O$_{\rm h}$ and Y$_{\rm
h}$; F$_4$ is the symmetry group of a regular solid in $\real^4$ with 24
3-dimensional octahedral faces; H$_4$ is the symmetry group of a regular solid
in $\real^4$ with 120 3-dimensional dodecahedral faces or, dually, of a
regular 600-sided solid with tetrahedral faces; the groups A$_3$, A$_4$, B$_4$
and D$_4$ are strictly related to permutation groups or semi-direct products
of permutation groups and sign change groups, as explained in \S 3.

Solid state physics is, therefore, a natural physical context where our
results can be exploited.  As an example of the use of the $\widehat P$-matrix
approach to the analysis of properties of an invariant function, and in
particular of the determination of the location of its stationary points and
of its absolute minimum, in this section we shall study a six-degree expansion
of a Landau thermodynamic potential $G(x;\pi,T)$, which depends on the vector
valued order parameter $x=(x_1,x_2,x_3)$, transforming according to the
fundamental representation of the group O$_h$.  At the end we shall specialize
our results to BaTiO$_3$ and determine its phase space using an oversimplified
expression for its free-energy.

\subsection{The orbit space of the group O$_{\rm h}$ and its stratification}
In the Coxeter notations used in the preceding sections, the fundamental
representation of the group O$_{\rm h}$ corresponds to the group B$_3$, for
which a MIB is specified in \eref{B3} and the corresponding $\widehat
P$-matrix in \eref{PB3} and \eref{LastRow}, with $d_1=6,\ d_2=4,\ d_3=2$.

To describe the geometry of the image $\bar{\cal S}$ of the orbit space of
O$_{\rm h}$, let us define the following auxiliary polynomial functions of
$p$:

\begin{eqnarray}\label{aux}
f_{1}(p) & = & -p_{2} + p_{3}^{2}\,, \nonumber \\
f_{2}(p) &=& -p_{1} + p_{3}^{3}\,,  \nonumber \\
f_{3}(p) &=& 3 p_{2} - p_{3}^{2}\,,  \nonumber \\
f_{4}(p) &=& 9 p_{1} - p_{3}^{3}\,,  \nonumber \\
f_{5}(p) &=& 2 p_{2} - p_{3}^{2}\,,  \nonumber \\
f_{6}(p) &=& 4 p_{1} - p_{3}^{3}\,,  \nonumber \\
f_{7}(p) &=& p_{3}^{3} - 3 p_{3} p_{2} + 2 p_{1} \,,  \nonumber \\
f_{8}(p) &=& - p_{3}^{6} + 9 p_{2} p_{3}^{4} - 8 p_{1} p_{3}^{3} -
21 p_{2}^{2}
p_{3}^{2} + 36 p_{1} p_{2} p_{3} + 3 p_{2}^{3} - 18 p_{1}^{2} \,.
\end{eqnarray}

The shape and primary stratification of $\bar{\cal S}$ can be immediately
deduced from an inspection of the condition

\eq{{\rm det}\,\widehat P(p) = f_7(p)f_8(p) = 0;}
the results obtained in this way are confirmed by a complete analysis of the
positivity and rank conditions on the matrix $\widehat P(p)$ defined in
\eref{PB3}.  The section $\Xi$ of the orbit space with the plane
$p_3=$const is plotted in figure 1, where the axes are labelled by
the adimensional variables

\eqll{u=p_2/p_3^2,\qquad v = p_1/p_3^3\,.}{eta}

The determination of the isotropy type stratification of $\bar{\cal S}$, that
is the determination of the orbit types of the different primary strata,
requires a sounder analysis:
for a convenient choice of the point $\bar p$ in each stratum,
one has first to find a solution $\bar x=({\bar x}_1, {\bar x}_2, {\bar x}_3)$
of the following equations\footnote{The numbers $({\bar x}_1, {\bar x}_2,
{\bar x}_3)$ are the
non-negative solutions of the real equation $\summ 03k a_kz^k=0$, with $a_0 =
(-\bar p_3^3 + 2\bar p_2\bar p_3 - 2\bar p_1)/6$ $(=-\bar x_1^2\bar x_2^2\bar
x_3^2)$, $a_1=(\bar p_3^2-\bar p_2)/2$ $(=\bar x_1^2\bar x_2^2+\bar x_2^2\bar
x_3^2+\bar x_3^2\bar x_1^2)$, $a_2=\bar p_3$ $(=\bar x_1^2 + \bar x_2^2 +
\bar x_3^2)$, $a_3=1$.}:

\eq{\label{x}
\left\{
\begin{array}{rcl}
x_{1}^{6} + x_{2}^{6} + x_{3}^{6}&=&{\bar p_1}   \,,\\
x_{1}^{4} + x_{2}^{4} + x_{3}^{4}&=&{\bar p_2}  \,, \\
x_{1}^{2} + x_{2}^{2} + x_{3}^{2}&=&{\bar p_3}  \,,
\end{array}
\right.}
and then, to determine the isotropy subgroup of O$_{\rm h}$ at $\bar x$, by
selecting the transformations of O$_{\rm h}$ which leave $\bar x$ invariant.

A solution $\bar x$ of \eref{x} for each stratum is easily found by noting
that

\begin{eqnarray}
{\rm det}\,\widehat P(p(x)) &=& f_7(p(x))f_8(p(x)) \nonumber\\
      &=& 36 x^2_1x^2_2x^2_3 (x^2_1 - x^2_2)^2 (x^2_2 - x^2_3)^2 (x^2_3 -
x^2_1)^2.
\end{eqnarray}

The results one obtains are reported in table 5.

\subsection{The absolute minimum of the potential. Phase transitions}
Let us now build the potential

\eq{G(x;\pi,T)=\widehat G(p(x);\pi ,T),}
as the most general sixth order O$_h$ invariant polynomial function of the
order parameters, the coefficients $c,A_i$ and $B_j$ being functions of the
pressure $\pi$ and the absolute temperature $T$:

\begin{equation}
\widehat G(p;\pi , T) = \frac{c}{2} p_{3} + \frac{1}{4} (A_{0} p_{3}^{2} +
A_{1} p_{2}) + \frac{1}{6} (B_{0} p_{3}^{3} + B_{1} p_{3} p_{2} + B_{2} p_{1})
\;.
\end{equation}

In order to assure stability of the system, we shall require that
$G(x;\pi,T)$ is bounded below.  This occurs if and only if the coefficient of
$p_3^3$ in the asymptotic form of $\widehat G$ for $p_3\rightarrow\infty$:

\eq{C_{\rm as}(u,v)=\frac{1}{6}
\left(B_{0} + B_{1} u + B_{2} v\right)}
is everywhere positive on $\Xi$, i.e.\ iff its minimum in $\Xi$ is $>0$.
Since the equation $C_{\rm as}(u,v)=$const defines straight lines in the
plane $(u,v)$ and $\bar{\cal S}$ is inscribed in a triangle with
vertices at the points representing the strata $\{{\rm C}_ {i{\rm v}}\}$,
$i=2,3,4$ (see figure 1), one easily realizes that $C_{\rm as} (u,v)$ is bound
to take on its absolute minimum at at least one of the vertices.  Therefore,
$\widehat G(p;\pi ,T)$ is bounded below if and only if the following
inequalities are satisfied:

\eq{\label{asy}
\left\{
\begin{array}{rcl}
C_{\rm as}(u,v)\big|_{\{{\rm C}_{2{\rm v}}\}} &=& (4B_0+2B_1 + B_2)/4\ >\
0\,,\\
C_{\rm as}(u,v)\big|_{\{{\rm C}_{3{\rm v}}\}} &=& (9B_0+3B_1 + B_2)/9\ >\
0\,,\\
C_{\rm as}(u,v)\big|_{\{{\rm C}_{4{\rm v}}\}} &=& B_0+B_1 +   B_2\ >\ 0\,.
\end{array}
\right.}

Being $\widehat G(p;\pi, T)$ a linear function of $(p_1,p_2)$, for fixed
$p_3$, the extremal points of $\widehat G(p; \pi ,T)$ lie necessarily on the
boudary of $\overline{\cal S}$ and can be determined \cite{021} using, for
instance, the standard method of Lagrange multipliers.  Denoting by $f_{A}=0,\
A\in{\cal I}^{(0)}_\alpha$ and $ f_{A}>0,\ A\in{\cal I}^{(+)}_\alpha$ the
algebraic relations defining the isotropy stratum $\Sigma_\alpha$, the
conditions one obtains can be written in the following form:

\begin{equation}
\left\{
\begin{array}{rcl}
{\displaystyle
\frac{\partial\widehat G}{\partial p_a}}&=& {\displaystyle\sum_{A\in {\cal
I}^{(0)}_\alpha} \lambda_A \frac{\partial f_A}{\partial p_A}\,,
\qquad a=1,2,\dots,q\,,} \\
f_{A} &=& 0 \,\qquad A\in {\cal I}^{(0)}_\alpha\,, \\
f_{A} &>&0 \,,\qquad A\in {\cal I}^{(+)}_\alpha\,,
\end{array}
\right.
\end{equation}
where the $\lambda$'s are Lagrange multipliers.

The determination of the absolute minimum and of its location(s) is made much
easier if one notes that
$\widehat G(p;\pi ,T)$ necessarily takes on its absolute minimum in some point
of one of the strata $\{{\rm O}_{{\rm h}}\}$, $\{{\rm C}_{i{\rm v}}\},\
i=2,3,4$\footnote{The possibility of a degenerate minimum crossing the strata
$\{{\rm C}_{2{\rm v}}\}$, $\{C'_{{\rm s}}\}$ and $\{{\rm C}_{4{\rm v}}\}$ can
be excluded by a direct check.}, like $C_{\rm as}(u,v)$ and for the same
reasons.

By indexing these strata in the following way:

\eq{\Sigma_0 = \{{\rm O}_{\rm h}\},\qquad \Sigma_1 = \{{\rm C}_{4{\rm
v}}\}, \qquad \Sigma_2 = \{{\rm C}_{2{\rm v}}\},\qquad \Sigma_3 =
\{{\rm C}_{3{\rm v}} \}\,,}
the relations determining the stationary points of the potential can be put in
a very compact form.

The relations defining the strata $\Sigma_k$, which can be read from table 5,
allow to express $p_1$ and $p_2$ in terms of $p_3$, so that we can define

\eq{\widehat G_k(p_3;\pi,T) =\widehat G(p;\pi,T)\big|_{\Sigma_k},\qquad k=0,
\dots ,3.}
Explicitly:

\begin{eqnarray}
\widehat G_0(0; \pi,T) &=& 0,\\
\widehat G_k(p_3;\pi,T) &=& p_3\left(\frac{c}{2} + \frac{a_kp_3}{4} +
\frac{b_kp_3^2}{6} \right),\qquad p_3>0,
\end{eqnarray}
where

\eq{
\begin{array}{ll}
\begin{array}{rcl}
a_k &=& A_0 + A_1\,k^{-1}\,, \\
b_k &=& B_0 + B_1\,k^{-1} + B_2\,k^{-2}\,,
\end{array}
\qquad k=1,2,3,
\end{array}
}
and, owing to \eref{asy}, $b_k\, >\,0$.

Now, it is trivial to realize that for $c\,<\,0$, each of the functions
$\widehat G_k(p_3;\pi,T),\ k=1,2,3$ has a local minimum, which is unique and
is located at $p_3=p_{3,k}$, where

\eqll{p_{3,k} = \frac{\left(-a_k +\sqrt{a_k^2 - 4b_kc}\right)}{2b_k}\,.}{p3k}
At $p_3=p_{3,k}$, the function $\widehat G_k(p_3;\pi, T)$ takes on the value:

\eqll{ \widehat G_k^{{\rm min}}(\pi, T) = \frac{1}{48b_k^2}\left(a_k -
\sqrt{a_k^2 - 4b_kc}\right)
\left(a_k^2 - 8b_kc - a_k\sqrt{a_k^2 - 4b_kc}\right)\,,\qquad k=1,2,3.}{G3k}

For $c\,\ge\,0$, only $\widehat G_1(p_3;\pi,T)$ has a local minimum, which is
unique and is located at $p_3=p_{3,1}$, with $p_{3,1}$ defined in \eref{p3k}
and $\widehat G_1^{{\rm min}}(\pi, T)$ defined in \eref{G3k}.

To determine the absolute minimum of $\widehat G(p;\pi, T)$ and the
phase space, it remains only to compare the values taken on by the
functions $\widehat G_k^{\rm min}(\pi, T)$, $k=0,\dots 3$, for $c\,<\,0$
and the functions $\widehat G_1^{{\rm min}}(\pi, T)$ and
$\widehat G_0^{{\rm min}}(\pi, T)=0$ in the plane $(\pi,T)$.

To be concrete, let us specialize our results to the case that $G(x;\pi,T)$
is an expansion in the order parameters of the free-energy of
BaTiO$_3$. Following Kim \cite{Kim}, we shall consider the following two
possibilities, in which a possible dependence on the pressure $\pi$ is
ignored and CGS units are used:

\begin{eqnarray}\label{ch1}
c     &=&  7.4 \cdot (T - 110)\, \cdot 10^{-5} \nonumber \\
B_{0} &=&  0 \,, \nonumber \\
A_{0} &=&  1.15 \cdot  10^{-12}\,,  \nonumber \\
A_{1} &=&  (-0.99 - 1.15) \cdot 10^{-12} \,,\nonumber \\
B_{1} &=&  0 \,,\nonumber \\
B_{2} &=&  0.249 \cdot 10^{-21} \,,
\end{eqnarray}

\begin{eqnarray}\label{ch2}
c     & = &  7.4 \cdot (T - 110) \cdot 10^{-5}\,, \nonumber \\
B_{0} & = &  0 \,,\nonumber \\
A_{0} & = &  12 \cdot 10^{-13}\,, \nonumber \\
A_{1} & = &  4 \cdot 4.5 \cdot (T - 175) \cdot 10^{-15} - 12 \cdot 10^{-13}
	     \,, \nonumber \\
B_{1} &=& 24 \cdot 10^{-23}\,, \nonumber \\
B_{2} &=&  30 \cdot 10^{-23}\,.
\end{eqnarray}
Then, using the data specified in \eref{ch1}, for $T>119.97\,^o{\rm C}$, the
free-energy takes on its absolute minimum at $\Sigma_0$; thus only the
disordered cubic phase $[{\rm O}_{\rm h}]$ is stable at these temperatures.
At $T=119.97\,^o{\rm C}$, the function $\widehat G_1^{{\rm min}}(T)$ vanishes,
so that the cubic and tetragonal phases coexist.  For $119.97\,^o{\rm
C}\,<\,T\,<\,14.77$ the absolute minimum sits on $\{{\rm C}_{4{\rm v}}\}$ and
the tetragonal phase $[{\rm C}_{4{\rm v}}]$ is stable.  At $T=14.77\,^o{\rm
C}$, the absolute minimum shifts to $\{{\rm C}_{2{\rm v}}\}$ and for
$14.77\,^o{\rm C}\,<\,T\,<\,-87.49$ the stable phase is the orthorhombic one.
At $T\,=\,-87.49$ the absolute minimum shifts to $\{{\rm C}_{3{\rm v}}\}$ and
for $T\,<\,-87.49$ the stable phase is the rhombohedral one.

These results, and the analogous ones obtained using the data specified in
\eref{ch2}, of \eref{ch2}, are resumed in tables 6 and 7.

If the free energy is expanded as a sufficiently high degree polynomial in the
order parameters $x$, and for a convenient choice of the coefficients as
functions of $T$, all the phases represented by the isotropy type strata of
the orbit space of O$_{\rm h}$ may become accessible (as stable phases) to the
system at convenient temperatures.  In particular, the sub-principal strata
require at least a 8th degree polynomial, while the principal stratum requires
at least a 12th degree polynomial.

\newpage

\section{Tables and table captions}

\begin{table}[h]
\caption{\label{tab1} Correspondence between classes of allowable solutions of
the canonical equation (labelled by the degree $d_1$) and finite coregular
generating groups, for $q=2$.}
\vspace{2em}
\begin{tabular}{c|cccccc}
\hline \hline
Group & $Z_2\otimes Z_2$ & A$_2$, I$_2(3)$ &  B$_2$, I$_2(4)$ & I$_2(5)$ &
G$_2$, I$_2(6)$ &  I$_2(m)$  \\
\hline
$d_1$ & 2               &  3    &    4  &   5   &    6   &   $m>6$ \\
\hline       \hline
\end{tabular}
\end{table}

\begin{table}[h]
\caption{\label{tab2} Correspondence between classes of
fundamental allowable solutions of the canonical equation,
degrees of the basic invariants  and finite coregular generating
groups, for $q=3$.}
\vspace{2em}
\begin{tabular}{c|cccc}
\hline   \hline
Group       & I$_2(m+1)\otimes $Z$_2$      & A$_3$, D$_3$ & B$_3$ & H$_3$\\
\hline
$(d_1,d_2)$ & $(m+1,2)$  & $(4,3)$ & $(6,4)$ & $(10,6)$ \\
\hline
Class       &   II$(m),\ m\in \natural_*$   &  III.1  &  III.2  &  III.3   \\
\hline \hline
\end{tabular}

\end{table}

\begin{table}[h]
\caption{\label{tab3} Correspondence between classes of irreducible
fundamental allowable solutions of the canonical equation,
degrees of the basic invariants  and finite coregular
irreducible generating groups, for $q=4$.}
\vspace{2em}
\begin{tabular}{c|ccccccccc}
\hline \hline
Group       & A$_4$     & D$_4$ & B$_4$ & F$_4$ &  H$_4$  \\
\hline
$(d_1,d_2, d_3)$ & $(5,4,3)$ & $(6,4,4)$  & $(8,6,4)$ & $(12,8,6)$ &
							  $(30,20,12)$  \\
\hline
Class       &   E1                     &  E2  &  E3  &  E4   &   E5    \\
\hline \hline
\end{tabular}
\end{table}

\begin{table}[h]
\caption{\label{tab4} Correspondence between classes of reducible
fundamental allowable solutions of the canonical equation,
degrees of the basic invariants  and finite coregular
reducible generating groups, for $q=4$.}
\vspace{2em}
\begin{tabular}{c|ccccccccc}
\hline \hline
Group & I$_2(j_1+1)\otimes $I$_2(j_2+1)$  & A$_3 \otimes $Z$_2$,
	D$_3 \otimes $Z$_2$ &  B$_3\otimes $Z$_2$ & H$_3\otimes $Z$_2$ \\
\hline
$(d_1,d_2, d_3)$ &  $((j_1+1),(j_2+1),2)$ & $(4,3,2)$ & $(6,4,2)$ & $(10,6,2)$
\\
\hline
Class &  A8$(j_1,j_2)\,,\ j_1\ge j_2\in\natural_*$ & B4(1) & C6(1) & D4(1) \\
\hline \hline
\end{tabular}
\end{table}

\begin{table}[h]
\caption{\label{tab5} Isotropy type strata of the orbit space of the
3-dimensional representation of the group O$_{\rm h}$.}
\vspace{2em}
\begin{tabular}{ccc}
\hline \hline
 Strata & Defining relations in $\real^q$  & Typical points in $\real^n$ \\
\hline
$\Sigma_0=\{O_{h}\}$ & $p_{1}= p_{2} = p_{3} = 0 $ & $\bar x_1 = \bar x_2 =
\bar x_3 = 0$ \\
\hline
$\Sigma_1=\{{\rm C}_{4{\rm v}}\} $ &  $ f_{1} = f_{2} = 0 < p_3$ & $\bar x_1
=1,\  \bar x_2 = \bar x_3 = 0$ \\
\hline
$\Sigma_3=\{{\rm C}_{3{\rm v}}\} $ &  $ f_{3} = f_{4} = 0 < p_3$ & $\bar x_1 =
\bar x_2 = \bar x_3 = 1$ \\
\hline
$\Sigma_2=\{{\rm C}_{2{\rm v}}\} $ &  $ f_{5} = f_{6} = 0 < p_3$ & $\bar x_1 =
0,\ \bar x_2 = \bar x_3 = 1$ \\
\hline
$\Sigma_4=\{{\rm C}_{s}\} $ &  $ f_{8} = 0 < f_{3},f_{7},p_3  $ & $\bar x_1 =
1,\ \bar x_2 = \bar x_3 = 2$\\
\hline
$\Sigma_5=\{C'_{s}\} $ &  $ f_{7} = 0 < f_{1}, f_{5},p_3 $ & $\bar x_1 = 0,\
\bar x_2 = 1, \bar x_3 = 2$\\
\hline
$\Sigma_{\rm p}=\{{\rm C}_{1}\} $ &  $ 0<f_{7} ,f_{8},p_3 $ & $\bar x_1 = 1,\
\bar x_2 = 2,\ \bar x_3 = 3 $\\
\hline \hline
\end{tabular}
\end{table}

\begin{table}[h]
\caption{\label{tab6} Stable phases of BaTiO$_3$ at different temperatures
with the first choice for the parameters involved in the definition
of the free-energy.}
\vspace{2em}
\begin{tabular}{ccc}
\hline \hline
\quad Phase\quad &\qquad & Range of temperatures\\
\hline
$[{\rm O}_{\rm h}]$=cubic & \qquad & $T\ >\ 119.97\,^o{\rm C}$ \\
\hline
$[{\rm O}_{\rm h}],\ [{\rm C}_{4{\rm v}}]$ & \qquad &
$T\ =\ T_{c_1}\ =\ 119.97\,^o{\rm C}$\\
\hline
$[{\rm C}_{4{\rm v}}]$=tetragonal & \qquad & $119.97\,^o{\rm C}\ >\ T\ >\
14.77\,^o{\rm C}$\\ \hline
$[{\rm C}_{4{\rm v}}],\ [{\rm C}_{2{\rm v}}] $ & \qquad &
$T\ =\ T_{c_2}\ =\ 14.77\,^o{\rm C}$\\
\hline
$[{\rm C}_{2{\rm v}}]=$orthorhombic  & \qquad & $14.77\,^o{\rm C}\ >\ T\ >
\ -87.49\,^o{\rm C}$\\
\hline
$[{\rm C}_{2{\rm v}}],\ [{\rm C}_{3{\rm v}}] $ & \qquad & $T\ =\ T_{c_3}\,=
\,-87.49\,^o{\rm C}$\\
\hline
$[{\rm C}_{3{\rm v}}]$=rhombohedral  & \qquad & $-87.49\,^o{\rm C}\ >\ T$\\
\hline \hline
\end{tabular}

\end{table}

\begin{table}[h]
\caption{\label{tab7} Stable phases of BaTiO$_3$ at different temperatures
with the second choice for the parameters involved in the definition of the
free-energy.}
\vspace{2em}
\begin{tabular}{cc}
\hline \hline
 Phase  &\qquad  Range of temperatures \\ \hline
[O$_{\rm h}$]\,=\,cubic & $T\,> 115.40\,{\rm ^oC}$ \\
\hline
[O$_{\rm h}$, [C$_{4{\rm v}}$] & $T\,=\, T_{{\rm c}_1}\,=\, 115.40\,{\rm ^oC}$
\\ \hline
[C$_{4{\rm v}}$]\,=\,tetragonal & $ 115.40\,{\rm ^oC}\,>\, T\,>\,
-233.52\,{\rm ^oC}$ \\
\hline
[C$_{4{\rm v}}$], [C$_{2{\rm v}}$] & $T\,=\, T_{{\rm c}_2}\,=\, -233.52\,{\rm
^oC}$ \\ \hline
[C$_{2{\rm v}}$]\,=\,orthorhombic & $ -233.52\,{\rm ^oC}\,>\,T $ \\
\hline \hline
\end{tabular}

\end{table}
\end{document}